\documentclass[twocolumn,longauthor,longbib,amsmath]{aastex702}
\usepackage[T1]{fontenc}
\usepackage[utf8]{inputenc}
\usepackage{amssymb,booktabs,array,calc}
\usepackage{needspace}
\usepackage{xurl}
\hypersetup{colorlinks=true,linkcolor=xlinkcolor,citecolor=xlinkcolor,urlcolor=xlinkcolor,pdftitle={Planetesimal Delivery to Giant Planet Circumplanetary Disks: Bulk Ice Enrichment and an Ice-rich Iapetus},pdfauthor={I. Mosqueira}}
\shorttitle{Ablation, capture, and satellite ice enrichment}
\shortauthors{Mosqueira}
\journalinfo{Draft version October 24, 2026}
\defcitealias{mosqueira2010}{MEC10}
\providecommand{\dodoi}[1]{doi:~\href{https://doi.org/#1}{\nolinkurl{#1}}}
\begin{document}
\title{Planetesimal Delivery to Giant Planet Circumplanetary Disks: Bulk Ice Enrichment and an Ice-rich Iapetus}
\author{I. Mosqueira}
\affiliation{San Jose State University, One Washington Square, San Jose, CA 95192,
USA}
\email[show]{Ignacio.Mosqueira@sjsu.edu}
\begin{abstract}
The regular satellites of Jupiter and Saturn are substantially enriched
in ice relative to the rock-rich solids of the outer Solar System. In addition,
the satellites interior to Titan and distant Iapetus are icier
than Titan itself. Collisional processes and subsequent migration can readily
alter compositions deep in the planetary potential well, but producing
a distant, ice-rich satellite with Iapetus's radius of about 700 km is a
challenge \citep{mosqueira2026a}. We show that differential ablation and capture
of icy and rocky fragments from differentiated outer Solar System
bodies can explain the bulk Jovian and Kronian satellite ice enrichment and supply the
substantial ice-rich inventory needed for Iapetus. For 1\,m--1\,km
fragments with the canonical size-distribution index \(p=3.5\) and a source containing
30\% ice by mass, the fixed radial distance Saturn calculation deposits material with ice
fractions of 0.66 (delivery of vapor-only rock) or 0.50 (with retained rock melt) at
20 Saturn radii and nearly pure ice at 60 Saturn radii. Following
three-dimensional trajectories over repeated passages through the
pressure-supported gas gives ice fractions of about 0.4 near 20 Saturn
radii and 0.9 near 60 Saturn radii. The matching Jupiter calculation
yields a similar bulk ice enrichment, with a corresponding compositional gradient across 
the Galilean satellite region. These delivery profiles supply ice-enriched material for
satellite formation. Yet satellitesimal inward migration subsequently modifies the material
incorporated into satellites and thus alters their bulk compositions. We briefly
discuss the implications for the PDS~70 planetary system.
\end{abstract}

\section{Introduction}\label{introduction}

The regular satellites of Jupiter and Saturn are enriched in ice
relative to the large, rock-rich bodies of the outer Solar System.
Titan, Ganymede, and Callisto illustrate this bulk compositional
difference (Section~\ref{observational-context}). Saturn's system also
contains satellites substantially icier than Titan, both interior and exterior
to its orbit.

\citet{mosqueira2026a} examined whether collisional processes within the
Kronian nebula could explain the observed satellite compositions.
Here we examine differential ablation and capture as an explanation
for the bulk ice enrichment of the Jovian and Kronian satellite systems
and the ice-rich composition of Iapetus. Planetesimal ablation of differentiated
icy and rocky fragments preferentially delivers ice to the circumplanetary disk,
as proposed by \citet[hereafter MEC10]{mosqueira2010}.
We model the disk ice-to-rock composition and mass delivered as a
function of radial distance in the circumplanetary disk, 
first for a single crossing at a fixed radial distance and
then for three-dimensional trajectories with repeated disk passages.
The resulting radial profile constrains the compositional gradient needed
to explain the marked ice enrichment of Iapetus. We likewise
apply the orbital calculation to Jupiter's circumplanetary disk, using a matching
fragment population and capture prescription, to characterize the corresponding bulk
ice enrichment of the Galilean satellite system.

As we discuss below, the PDS~70 planetary system provides an additional 
setting for these processes, with two
neighboring giant planets and a persistent circumplanetary dust
reservoir around PDS~70~c. The disk extends to about \(0.1R_H\),
where \(R_H\) is the planetary Hill radius \citep{mosqueira2026pds}. Sustained
interactions between incoming solids and circumplanetary gas make
ablation and capture processes natural contributors to satellite compositions
in the PDS~70 system as well, although the relevant parameters
are poorly constrained at present. Section~\ref{pds70-analogy}
develops the analogy with the Iapetus region and the possibility of
observing ice enrichment in the outer disk of PDS~70~c.

We stress that satellitesimals drift in and growing satellites also 
migrate inward, so each
satellite incorporates material supplied over a range of radii. 
The modeled ablation delivery
profile therefore provides the initial disk compositional 
constraints applicable to satellite formation.
Titan, the dominant repository of satellite material, anchors
the bulk composition of the Kronian system.
Iapetus probes the supply of ice-rich material in the outer disk
\citep{mosqueira2005}. Collisional processes together with migration can modify
the compositions of the rings and inner moons deep in the planetary
potential well.

For specificity,
we adopt an end-member source of separate icy and rocky fragments with an
aggregate 30/70 ice--rock mass ratio (Section~\ref{heterogeneous-source}). Section~\ref{relation-to-the-earlier-papers}
summarizes the changes from \citetalias{mosqueira2010}. Sections~\ref{disk-and-incoming-fragments}--\ref{from-fragment-efficiencies-to-a-disk-composition} specify the model,
Sections~\ref{calculated-delivery-and-compositional-gradients}--\ref{orbital-delivery-and-the-compositional-gradient} give the results, and Sections~\ref{discussion}--\ref{conclusions} discuss and summarize
them. We also discuss the implications of migration for the final satellite compositions.

\subsection{Bulk satellite ice enrichment and the composition of
Iapetus}\label{observational-context}

\begin{figure}[t!]
\centering
\includegraphics[width=1\linewidth,keepaspectratio,alt={Rock-plus-metal mass fraction versus volume-equivalent mean radius for bodies at least as large as Quaoar, with the complementary water-rich fraction on the right. Titan, Ganymede, and Callisto lie below the source reference line; Iapetus lies farther below them at about 20 percent rock. The dotted line marks the adopted 70 percent refractory source. Published models: Titan from Sohl et al. 2003; Ganymede and Callisto from Kuskov and Kronrod 2001 and 2005; Pluto and Charon from McKinnon et al. 2017; Triton from Cioria and Mitri 2022. Compact reference calculations from Appendix~\ref{a.7.-size-data-and-the-compact-reference-calculations} are shown for Quaoar, Iapetus, Haumea, and Eris.}]{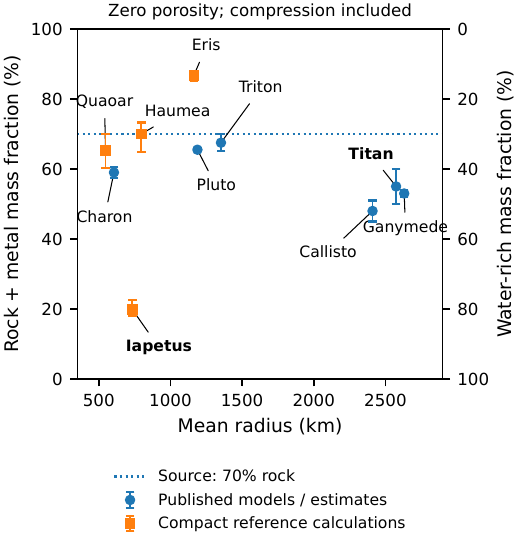}
\caption{Rock-plus-metal mass fraction versus volume-equivalent mean
radius for bodies at least as large as Quaoar. The right axis gives the
complementary water-rich fraction. Porosity is neglected and compression
is included. \textbf{Circles:} published interior models or
estimates: Titan, \citet{sohl2003}; Ganymede and Callisto,
\citet{kuskov2001,kuskov2005}; Pluto and Charon, \citet{mckinnon2017};
Triton, \citet[Introduction]{cioria2022}. \textbf{Squares:} compact
reference calculations (Appendix~\ref{a.7.-size-data-and-the-compact-reference-calculations}) using the material constants of
\citet{fortes2012}. Vertical bars span the ranges of those models or
sensitivity calculations, with the definitions given in Table~\ref{tab:A1}.
Symbols mark range midpoints. The dotted line is the source's 70\%
refractory fraction. Table~\ref{tab:A1} lists all
values.}\label{fig:observational-context}
\end{figure}

Figure~\ref{fig:observational-context} compares the inferred refractory (rock-plus-metal) mass
fractions of outer Solar System bodies at least as large as Quaoar\footnote{The densities of
smaller bodies can be significantly affected by macroporosity.}.
For Titan, Ganymede, and Callisto we adopt published interior models
that include high-pressure water phases. Appendix~\ref{a.4.-composition-references-and-the-saturn-disk-overlays} and Table~\ref{tab:A1} give
the sources and values.

Titan illustrates why compression matters.
\citet{sohl2003} obtain 50--60\% anhydrous rock plus iron in models that
contain high-pressure ice and water-rich liquid layers. Treating all of
Titan's water as ice of density \(0.92\ \mathrm{g\,cm^{-3}}\) would
instead give about 70\% rock. Ganymede (52--54\%
refractory; \citealp{kuskov2001}) and Callisto (45--51\%;
\citealp{kuskov2005}) are similarly ice-rich. By contrast, Triton has
65--70\% rock plus metal \citep{cioria2022}, the reference interior of
Pluto has 66\% \citep{mckinnon2017}, and the compact reference
calculations give about 65\% for Quaoar, 70\% for Haumea, and 87\% for
Eris. The large regular satellites are therefore enriched in ice
relative to the large, rock-rich bodies of the outer Solar System.

In these models Callisto is also icier than Ganymede, which leaves open
an outward gradient in the Jovian system, although the Titan--Iapetus
contrast is more pronounced. Saturn's full compositional pattern
includes ice-rich moons on both sides of Titan. Interior examples
are Tethys and Rhea, whose mean densities are about 1.0 and
\(1.2\ \mathrm{g\,cm^{-3}}\), respectively \citep{jpl2026a}.
Tethys's density is close to that of water ice; Rhea's ice-dominated
composition is discussed by \citet{anderson2007}.

Iapetus's composition stands in stark contrast to that of Quaoar. 
Quaoar is smaller than Iapetus
(volume-equivalent radius about 550 km against 730 km) yet it is much denser:
\(1.76\pm0.11\) against \(1.089\pm0.013\ \mathrm{g\,cm^{-3}}\)
\citep{margoti2026,jpl2026a}. The compact reference calculations give
about 65\% and 20\% refractory material, respectively. Attributing the
low density of Iapetus to porosity would require explaining how that
porosity arose and persisted in the larger body. Let \(P_c\) denote
central pressure, \(G\) the gravitational constant, and \(\rho_I\) and
\(R_I\) Iapetus's mean density and radius. A uniform-density estimate
gives

\[
P_c\simeq\frac{2\pi}{3}G\rho_I^2R_I^2\simeq8.9\times10^7\ \mathrm{Pa}.
\]

Let \(\rho_r\) and \(\rho_i\) denote the rock and ice densities and
\(\rho_{\rm compact}\) their mixture density without voids. A compact
30/70 ice--rock mixture at low pressure with \(\rho_r=3.4\) and
\(\rho_i=0.92\ \mathrm{g\,cm^{-3}}\) has
\(\rho_{\rm compact}\simeq1.9\ \mathrm{g\,cm^{-3}}\). Reproducing
Iapetus's density would require a whole-body void-volume fraction
\(\phi_{\rm required}\) given by

\[
\phi_{\rm required}=1-\frac{\rho_I}{\rho_{\rm compact}}\simeq0.42,
\]

not merely a porous surface layer. Cold-ice compaction experiments and
the self-gravity and thermal arguments make such bulk porosity
difficult to sustain \citep{durham2005}.

\subsection{A heterogeneous source of icy and rocky planetesimal
fragments}\label{heterogeneous-source}

We adopt an aggregate source of icy and rocky planetesimal fragments
that contains 30\% ice and 70\% rock by mass, following
\citetalias{mosqueira2010}\footnote{The actual
compositions of the fragments may be more complex than those in our simple two-component model,
resulting in more rock delivery.
On the other hand, inward migration can enrich the inner disk with 
ice-rich material from the 
outer disk.}. 
The density dichotomy of
trans-Neptunian objects \citep{canas2024} and the JWST
surface-compositional groups \citep{pinilla2025} both point to
heterogeneous primordial environments.
A population of icy and rocky fragments can arise from differentiation followed
by collisional disruption. 
Early-forming heliocentric parent bodies differentiated through heating
from the decay of their initial \(^{26}\)Al inventory.

The Haumea family may illustrate compositional selection through release of icy
material from a differentiated progenitor \citep{brown2007,brown2012}. Collision and
merger calculations preferentially release mantle material while
retaining most of the rock in a large remnant
\citep{leinhardt2010,proudfoot2022}.

\subsection{Relation to the earlier
papers}\label{relation-to-the-earlier-papers}

\citet{mosqueira2005} proposed preferential ablation followed by local
formation. \citetalias{mosqueira2010} developed this into a quantitative model that couples
mass loss and drag, integrates separate ice and rock size distributions,
and calculates radial composition and delivered mass.

In this study, we first introduce a revised surface mass-loss law and an
orbital calculation of repeated disk passages. In
\citetalias{mosqueira2010}, ice loss began at melting (273 K) or
vaporization (648 K), and rock followed an effective melting law near
1800 K; \citetalias{mosqueira2010} anticipated additional outer ice
delivery by sublimation below those thresholds. Here vapor pressure sets
the mass loss, and the surface temperature follows from radiation and
removal energies \citep{dangelo2015}. For rock, vapor and shed melt share
one energy budget: the vapor energy is deducted before the remaining
power is assigned to melting. The melting temperature and effective
melting energy are unchanged. Section~\ref{rock-melting-the-revised-mass-loss-prescription} specifies the revised
law, and Appendix~\ref{a.6.-comparison-of-the-old-and-revised-melting-laws} compares it with \citetalias{mosqueira2010}.

Our two rock treatments hold the ice law fixed and compare rock vapor
loss alone (\emph{vapor-only}) with vapor loss plus retained melt
shedding (\emph{retained melt}). \citetalias{mosqueira2010}'s curves 
instead varied the ice
treatment while keeping its effective rock-melting law.

This study also removes the fixed radial distance, single-crossing restriction.
We follow three-dimensional trajectories through rotating gas, propagate
survivors from their actual exit states, and distinguish first entry
from the return and capture of previously processed objects. This
measures a contribution deferred in the earlier work, and the radial
composition curves use actual deposition radii, including
delivery on later passages. 

Coupled orbital evolution and ablation have
been studied in three-dimensional gas models of planetary accretion
by \citet{dangelo2015}. We apply orbital continuation to separate icy and
rocky populations and determine how returning, already processed
fragments change the delivered mixture and its radial gradient in the
circumplanetary disk.

\section{Disk and incoming
fragments}\label{disk-and-incoming-fragments}

\subsection{A fixed, separated source
population}\label{a-fixed-separated-source-population}

Let \(j=i,r\) label ice and rock, \(N_j\) the number of fragments of
component \(j\), and \(s_0\) their initial radius. The number
normalization is \(A_j\), the size-distribution index is \(p\), and
the lower and upper radius limits are \(s_{\min}\) and \(s_{\max}\). We
use

\begin{equation}
\frac{dN_j}{ds_0}=A_j s_0^{-p},\qquad
s_{\min}\leq s_0\leq s_{\max}.
\label{eq:legacy-1}
\end{equation}

Both components share this radius distribution; their amplitudes are set
by the prescribed incoming component masses. The reference calculation
uses \(p=3.5\), \(s_{\min}=1\) m, \(s_{\max}=10^3\) m, and source ice
fraction \(f_0=0.30\), with fragment densities \(\rho_i=920\) and
\(\rho_r=3400\ \mathrm{kg\,m^{-3}}\). Mass weighting is proportional to
\(s_0^{4-p}\,d\ln s_0\), not to fragment number.

Because the densities differ, so do the number amplitudes. Let \(M_j\)
be the incoming mass of component \(j\). For the same index and size
limits,

\begin{equation}
\frac{A_i}{A_r}=\frac{f_0}{1-f_0}\frac{\rho_r}{\rho_i}\simeq1.6,
\qquad
\frac{dM_j}{d\ln s_0}\propto s_0^{4-p}.
\label{eq:legacy-1a}
\end{equation}

Thus the component mass ratio is 30/70 even though the number amplitudes
differ. At \(p=3.5\), the intervals 1--10, 10--100, and 100--1000 m
contain 7.1\%, 22\%, and 71\% of each component's incoming mass. The
efficiency integral includes all three intervals. We also calculate
\(p=3\) and \(p=4\) with the same two components, size bounds, disk, and
heat-transfer laws; only the mass weighting changes.

The 30/70 component ratio, separate-fragment premise, and \(p=3.5\)
comparison follow \citetalias{mosqueira2010}. Our size-distribution index \(p\) is \citetalias{mosqueira2010}'s \(q\); their separate
mass-distribution exponent is \((q+2)/3\). The upper cutoff here is 1
km, rather than the 50 km upper cutoff of \citetalias{mosqueira2010}'s
Saturn example. All efficiencies and supplied-mass normalizations below
refer to this stated 1--1000 m population.

\subsection{Reference circumplanetary
disk}\label{reference-circumplanetary-disk}

Let \(R\) be cylindrical distance from Saturn's spin axis, \(R_S\) the
adopted Saturn radius, and \(M_S\) Saturn's mass. Write \(X=R/R_S\),
with \(R_S=6.0\times10^7\) m and
\(GM_S=3.79\times10^{16}\ \mathrm{m^3\,s^{-2}}\). The inner and outer
reference gas-column branches, \(\Sigma_{\rm in}\) and
\(\Sigma_{\rm out}\), are

\begin{equation}
\Sigma_{\rm in}=2\times10^4\frac{20}{X},\qquad
\Sigma_{\rm out}=2\times10^3\left(\frac{58}{X}\right)^2
\quad [\mathrm{g\,cm^{-2}}].
\label{eq:legacy-2}
\end{equation}

These are \citetalias{mosqueira2010}'s Saturn branches, joined between
\(20R_S\) and \(58R_S\) by the interpolation in Appendix~\ref{a.1.-continuous-disk-interpolation}. The gas
temperature is \(T_g=\max(2000/X,90)\) K. Let \(z\) be height above the
midplane, \(\rho_g\) gas density, \(\Sigma_g\) the interpolated full gas
column, and \(H\) the vertical scale height. With Keplerian frequency
\(\Omega_K=(GM_S/R^3)^{1/2}\), sound speed \(c_s\), Boltzmann constant
\(k_B\), and hydrogen-atom mass \(m_H\), the vertically isothermal disk
is

\begin{equation}
\begin{gathered}
\rho_g(R,z)=\frac{\Sigma_g(R)}{\sqrt{2\pi}H}
\exp\!\left(-\frac{z^2}{2H^2}\right),\\
H=\frac{c_s}{\Omega_K},\qquad
c_s^2=\frac{k_BT_g}{2.34m_H}.
\end{gathered}
\label{eq:legacy-3}
\end{equation}

The factor 2.34 is the adopted mean gas-particle mass in units of
\(m_H\).

The fixed radial distance grid covers \(8R_S\)--\(100R_S\); the orbital
calculation extends the gas inward to \(2R_S\) (Section~\ref{orbital-continuation-and-repeated-disk-passages}). The adopted outer
boundary is \(100R_S\), or about \(0.1R_H\). The gas structure is held fixed during
each encounter. The radiation
background is set to \(T_g\) for the reference thermal calculation.
Table~\ref{tab:1} summarizes the reference crossing inputs.

\begin{table*}[t!]
\centering
\caption{Inputs to the normal-crossing calculation. The two radii represent the
Titan and Iapetus regions at the precision appropriate to this reference
model. The actual surface temperature and speed evolve during the
crossing.\label{tab:1}}
\small\setlength{\tabcolsep}{4pt}\renewcommand{\arraystretch}{1.15}
\begin{tabular}{@{}>{\raggedright\arraybackslash}p{(\linewidth - 4\tabcolsep)*\real{0.54000000}}>{\raggedright\arraybackslash}p{(\linewidth - 4\tabcolsep)*\real{0.23000000}}>{\raggedright\arraybackslash}p{(\linewidth - 4\tabcolsep)*\real{0.23000000}}@{}}
\toprule
Quantity & Inner reference, \(20R_S\) & Outer reference, \(60R_S\)\\
\midrule
Full Gaussian column, \(\Sigma_g\) (\(\mathrm{g\,cm^{-2}}\)) & \(2.0\times10^4\) & \(1.9\times10^3\)\\
Gas and background temperature (K) & 100 & 90\\
Midplane density (\(\mathrm{g\,cm^{-3}}\)) & \(6.3\times10^{-7}\) & \(1.2\times10^{-8}\)\\
Imposed initial crossing speed (\(\mathrm{km\,s^{-1}}\)) & 8.3 & 5.2\\
\bottomrule
\end{tabular}
\end{table*}

\section{Ablation and capture of an individual
fragment}\label{ablation-and-capture-of-an-individual-fragment}

\subsection{Coupled motion and changing
radius}\label{coupled-motion-and-changing-radius}

We retain \citetalias{mosqueira2010}'s local, perpendicular crossing
geometry. The fragment begins at \(z=-H\) and is followed until it
emerges at \(+H\), is locally confined, or is effectively ablated. The
path between these boundaries contains
\(\operatorname{erf}(1/\sqrt{2})\simeq0.68\) of the full Gaussian column
in equation~\eqref{eq:legacy-3}, defining the reference crossing column.

Let \(s\) be the instantaneous fragment radius, \(u\) its normal
gas-relative velocity, and \(\rho_j\) its material density. The vapor
and shed-melt mass fluxes per unit surface area are \(J_{v,j}\) and
\(J_{m,j}\). We use spherical fragments, isotropic mass loss, and drag
coefficient \(C_D=0.44\). Hereafter, trajectory time \(t\) is measured
from entry, and overdots denote time derivatives. At fixed \(R\), the
one-dimensional equations are

\begin{equation}
\begin{aligned}
\dot z&=u,\\
\dot u&=-\frac{3C_D\rho_g}{8\rho_js}\,u|u|-\Omega_K^2z,\\
\dot s&=-\frac{J_{v,j}+J_{m,j}}{\rho_j}.
\end{aligned}
\label{eq:legacy-4}
\end{equation}

Let \(u_0\) be the imposed entry speed and \(\Theta\) the dimensionless
focusing parameter, defined by \(v_\infty^2=GM_S/(R_S\Theta)\) for
asymptotic planet-centered speed \(v_\infty\). Then

\begin{equation}
u_0^2=\frac{GM_S}{R_S\Theta}+\frac{2GM_S}{R},\qquad \Theta=100.
\label{eq:legacy-5}
\end{equation}

The whole speed in equation~\eqref{eq:legacy-5} is assigned to the
normal gas-relative velocity: the crossing is vertical at fixed radial distance,
not a three-dimensional orbit in rotating gas.

We follow intact parents as they ablate and use a strength diagnostic
to assess the ram pressure along their trajectories. For an
icy fragment with an initial radius of 1 km at \(10R_S\), the calculated peak ram
pressure \(\rho_g u^2/2\) is \(1.6\times10^5\) Pa. Along that
trajectory, its maximum ratio to the nominal ice strength
\(10^5(s/1\,\mathrm{km})^{-1/2}\) Pa adopted by \citet{dangelo2015} is
1.5; at \(11R_S\) it is 1.2. Under that prescription self-gravity would
not prevent breakup, although stronger compact ice could survive. The
same screen stays below threshold at \(20R_S\) and \(60R_S\). These strength diagnostics accompany the intact-parent inventories in
Section~\ref{sufficient-material-at-the-relevant-locations}; the
calculation retains the parent until ablation or its dynamical endpoint.

\subsection{Surface heating and vapor
loss}\label{surface-heating-and-vapor-loss}

With heat-transfer coefficient \(C_H\), the aerodynamic power is
\(\dot Q_{\rm aero}=C_H\rho_g\pi s^2|u|^3/2\). Its flux per unit
spherical surface area, \(q_{\rm aero}\), is

\begin{equation}
q_{\rm aero}=\frac{C_H\rho_g|u|^3}{8},\qquad C_H=0.10.
\label{eq:legacy-6}
\end{equation}

With this convention, \citeauthor{dangelo2015}'s
\citeyearpar{dangelo2015} heating prescription corresponds to
\(C_H=C_D/4\), or 0.11 at \(C_D=0.44\). We retain the nearby value 0.10
used in \citetalias{mosqueira2010}. Heating decreases as the fragment
decelerates; a fixed entry-speed fluence is not applied after drag has
slowed it.

Let \(T_s\) be surface temperature, \(P_j(T_s)\) the component vapor
pressure, \(\mathcal M_j\) its effective molar mass, and \(R_g\) the
molar gas constant. For an exposed surface with negligible adjacent
vapor partial pressure and unit evaporation coefficient,

\begin{equation}
J_{v,j}(T_s)=P_j(T_s)
\sqrt{\frac{\mathcal M_j}{2\pi R_gT_s}}.
\label{eq:legacy-7}
\end{equation}

Let \(Q_{v,j}\) and \(Q_{m,j}\) be the effective energy costs per unit
mass removed as vapor and melt. With emissivity \(\epsilon\) and
Stefan--Boltzmann constant \(\sigma_{\rm SB}\), \(T_s\) is solved
simultaneously from

\begin{equation}
\begin{aligned}
q_{\rm aero}+\epsilon\sigma_{\rm SB}T_g^4
&=\epsilon\sigma_{\rm SB}T_s^4\\
&\quad+Q_{v,j}(T_s)J_{v,j}+Q_{m,j}J_{m,j},\\
\epsilon&=0.5.
\end{aligned}
\label{eq:legacy-8}
\end{equation}

The steady receding-surface approximation assigns vapor an effective
removal enthalpy \(Q_{v,j}\), including phase change and sensible
heating of fresh material initially at 100 K. Appendix~\ref{a.2.-material-laws-and-effective-removal-energies} gives the
material laws. No separate internal-energy equation or additional
conductive-storage sink is imposed; transient internal heating is not
resolved.

The surface solution permits ice loss below melting and determines the
refractory vapor flux continuously. Both enter the radius evolution in
equation~\eqref{eq:legacy-4}.

\subsection{Rock melting: the revised mass-loss
prescription}\label{rock-melting-the-revised-mass-loss-prescription}

We calculate two rock treatments with identical disk and source inputs.
The vapor-only treatment sets \(J_{m,r}=0\) and solves equations~\eqref{eq:legacy-7}--\eqref{eq:legacy-8} for the surface temperature and refractory vapor flux. The
retained-melt treatment permits efficient shedding of molten rock, which
is credited to the local deposit (Section~\ref{local-capture-and-the-deposited-mass}). It retains
\citetalias{mosqueira2010}'s melting temperature, \(T_m=1800\) K, and
the cold-material melting energy in its Table 1,
\(Q_{m,r}=3.04\times10^6\ \mathrm{J\,kg^{-1}}\)
(\(3.04\times10^{10}\ \mathrm{erg\,g^{-1}}\)). Neither constant is
adjusted in this comparison.

Below the power needed to maintain melting and the associated vapor
loss, the vapor-only equilibrium applies. Once that power is exceeded,
the surface is held at \(T_m\), vapor loss is \(J_{v,r}(T_m)\), and the
remaining energy removes melt:

\begin{equation}
J_{m,r}=\frac{
q_{\rm aero}-\epsilon\sigma_{\rm SB}(T_m^4-T_g^4)
-Q_{v,r}(T_m)J_{v,r}(T_m)}{Q_{m,r}}.
\label{eq:legacy-9}
\end{equation}

Equation~\eqref{eq:legacy-9} is used only when its numerator is positive. The vapor and
melt fluxes remove material from the same fragment, and
\(\dot s=-(J_{v,r}+J_{m,r})/\rho_r\) updates the radius used in the drag
acceleration. The melting energy includes warming fresh rock from 100 K
and fusing it; vaporized material pays the distinct
cold-material-to-vapor energy \(Q_{v,r}\). Fusion is not charged again
on top of either stated removal energy.

Unlike \citetalias{mosqueira2010}'s single effective melt-ablation flux,
equation~\eqref{eq:legacy-9} deducts the vapor energy before assigning the remainder to
shed melt. It reduces to the earlier energy-limited form when vapor loss
is negligible and the radiation convention and removal energy are
matched. Appendix~\ref{a.6.-comparison-of-the-old-and-revised-melting-laws} gives the old law, the algebraic comparison, and
the activation-flux calculation.

We use efficient shedding: excess power processes and releases fresh
material while the surface stays at the melting temperature.
Meteoroid-ablation models explicitly couple vaporization with
aerodynamic removal of a molten surface layer \citep{dias2020}.
The effective removal law specifies melt loss from the parent; the
retention rule below specifies its contribution to the disk.

\subsection{Local confinement and the deposited
mass}\label{local-capture-and-the-deposited-mass}

Local confinement is kept distinct from a remnant merely becoming
gravitationally bound to Saturn. \citetalias{mosqueira2010} excluded
emerging remnants from the local deposited mass because their later
delivery need not occur at the crossing radius. We retain that
definition for the fixed radial distance reference calculation, then follow
emerging remnants dynamically in the orbital extension below.

For equation~\eqref{eq:legacy-4}, define the vertical specific energy \(E_z\) by

\begin{equation}
E_z=\frac12\left(u^2+\Omega_K^2z^2\right),\qquad
\dot E_z=-\frac{3C_D\rho_g}{8\rho_js}|u|^3\leq0.
\label{eq:legacy-10}
\end{equation}

Once \(E_z\leq\Omega_K^2H^2/2\) inside the interval, neither boundary is
accessible even without further drag. We register local confinement and
credit the remaining parent mass as confined. This sufficient stopping
condition avoids following repeated vertical oscillations; it is not
Saturn-centered escape energy.

Let \(m_{0,j}\) and \(m_{\rm rem,j}\) be the initial and surviving
fragment masses. The indicator \(\mathbf 1_{\rm confined}\) is one for a
confined remnant and zero otherwise. The fractions \(\mathcal A_j\),
\(\mathcal B_j\), and \(\mathcal C_j\) record vapor loss, shed melt, and
confined remnant mass, respectively:

\begin{equation}
\begin{aligned}
\mathcal A_j&=\frac{1}{m_{0,j}}\int4\pi s^2J_{v,j}\,dt,\\
\mathcal B_j&=\frac{1}{m_{0,j}}\int4\pi s^2J_{m,j}\,dt,\\
\mathcal C_j&=\frac{m_{\rm rem,j}}{m_{0,j}}\,\mathbf 1_{\rm confined}.
\end{aligned}
\label{eq:legacy-11}
\end{equation}

Let \(D_j\) be the locally deposited fraction and \(\mathcal E_j\) the
emerging-remnant fraction. Then

\begin{equation}
D_j(R,s_0)=\mathcal A_j+\mathcal B_j+\mathcal C_j,
\qquad
D_j+\mathcal E_j=1.
\label{eq:legacy-12}
\end{equation}

Only surviving mass enters \(\mathcal C_j\), so material previously
ablated is not counted again. Later processing of a confined remnant
does not change the total local inventory.

\textbf{Retention of shed melt.} We assume complete local retention of
both vapor and shed melt. Vapor joins the disk gas, with water available
after recondensation. Throughout the delivery calculation, the reported
ice fraction is the mass fraction of the delivered water component,
including vapor, irrespective of its phase at deposition. Molten rock
leaves the parent as droplets small
enough to be stopped within the remaining gas column. Let
\(\dot m_{\rm dep,m}\) and \(\dot m_{\rm shed,m}\) denote the deposited
and shed melt mass rates. The corresponding choice is

\begin{equation}
\dot m_{\rm dep,m}=\dot m_{\rm shed,m}=4\pi s^2J_{m,r}.
\label{eq:legacy-12a}
\end{equation}

This effective retention rule credits shed melt to \(\mathcal B_r\),
whereas \(\mathcal C_r\) is surviving parent mass satisfying
equation~\eqref{eq:legacy-10}. This retains the effective delivery
assumption of \citetalias{mosqueira2010}. The remaining mass of
throughgoing parents is not credited to the local annulus; its later
delivery is calculated separately rather than assigned to the entry
radius.

\subsection{Orbital dynamics and pressure-supported gas}\label{orbital-continuation-and-repeated-disk-passages}

To determine which remnants return, we extend the calculation to
three-dimensional trajectories in the circular Hill approximation,
following the ballistic approach used for PDS~70 by
\citet{mosqueira2026pds}. The disk and Saturn's heliocentric orbit are
coplanar. Let \(M_\odot\) be the solar mass and \(a_p=9.58\) au Saturn's
heliocentric orbital radius. The Saturn-centered frame rotates at
angular frequency \(n_p=(GM_\odot/a_p^3)^{1/2}\). Its Cartesian
coordinates have \(x\) toward the Sun, \(y\) completing a right-handed
frame, and \(z\) normal to the disk. Position is
\(\boldsymbol r=(x,y,z)\), spherical distance is \(r=|\boldsymbol r|\),
and cylindrical distance is \(R=(x^2+y^2)^{1/2}\). Let
\(\boldsymbol v_{\rm rot}\) and \(\boldsymbol v_{g,\rm rot}\) be the
fragment and gas velocities in this frame,
\(\boldsymbol w=\boldsymbol v_{\rm rot}-\boldsymbol v_{g,\rm rot}\)
their relative velocity, and \(\boldsymbol a_D\) the drag acceleration,
with Cartesian components \(a_{D,x}\), \(a_{D,y}\), and \(a_{D,z}\). The
dimensional equations are:

\begin{equation}
\begin{aligned}
\ddot x&=2n_p\dot y+3n_p^2x-GM_Sx/r^3+a_{D,x},\\
\ddot y&=-2n_p\dot x-GM_Sy/r^3+a_{D,y},\\
\ddot z&=-n_p^2z-GM_Sz/r^3+a_{D,z},\\
\boldsymbol a_D&=-\frac{3C_D\rho_g}{8\rho_j s}
|\boldsymbol w|\boldsymbol w.
\end{aligned}
\label{eq:legacy-12b}
\end{equation}

The gas rotation includes radial pressure support. Let
\(\rho_{g,0}(R)=\rho_g(R,0)\) be the midplane density,
\(P_0=\rho_{g,0}c_s^2\) the midplane pressure, and \(\beta\) its
negative logarithmic radial slope. Midplane radial force balance gives
the inertial azimuthal speed \(v_{g,\phi}\):

\begin{equation}
\begin{aligned}
\beta(R)&=-\frac{d\ln P_0}{d\ln R},\\
v_{g,\phi}^2&=\frac{GM_S}{R}+\frac{R}{\rho_{g,0}}\frac{dP_0}{dR}\\
&=\frac{GM_S}{R}-\beta c_s^2.
\end{aligned}
\label{eq:legacy-12c}
\end{equation}

With \(\boldsymbol e_\phi\) the prograde azimuthal unit vector, the
rotating-frame gas velocity is

\[
\boldsymbol v_{g,\rm rot}=(v_{g,\phi}-n_pR)\boldsymbol e_\phi.
\]

The pressure derivative follows the smooth surface-density interpolation
and the appropriate branch of the temperature floor, and the squared
speed is positive throughout the gas domain. We extend this midplane
rotation law vertically at fixed cylindrical radius, using steady,
azimuthal gas flow.
Appendix~\ref{a.10.-gas-rotation-and-physical-boundary-sensitivities}
gives a Keplerian-gas numerical control.

Drag and heating use the same Gaussian density and surface-energy
balance, evaluated at the projectile's instantaneous position with
\(|\boldsymbol w|\) in place of \(|u|\). Thermal recession updates the
radius, and hence the drag, throughout every encounter. Drag and
ablation act for \(2\leq R/R_S\leq100\) and \(|z|\leq H(R)\), with the
prescribed inner surface-density and temperature laws extended to
\(R=2R_S\). The outer gas edge is about \(0.1R_H\). The vertical boundary retains the reference column truncation while
permitting radial motion and repeated crossings. Outside this gas
domain, the remnant follows the ballistic Hill equations until it
reaches an absorbing spherical boundary at \(r=2R_S\) or the outer exit
boundary. The fragment receives no velocity or size reset at a
boundary.

Subscript 0 denotes the entry state: \(\boldsymbol r_0\) is its
position, \(r_0=|\boldsymbol r_0|\) its spherical distance, and \(R_0\)
its cylindrical radius. Controlled trajectory examples begin at
\(R_0=20R_S\) or \(60R_S\), on the lower gas surface. The imposed
planet-centered inertial speed is
\(v_0=[v_\infty^2+2GM_S/r_0]^{1/2}\), with the same
\(v_\infty^2=GM_S/(R_S\Theta)\) and \(\Theta=100\). Let
\(v_R,v_\phi,v_z\) denote the inertial radial, azimuthal, and vertical
velocity components. The displayed examples enter vertically, with
\((v_R,v_\phi,v_z)_0=(0,0,v_0)\), from
\(\boldsymbol r_0=(R_0,0,-H(R_0))\). Let \(\boldsymbol e_z\) be the unit
vector normal to the disk and \(\boldsymbol n_p=n_p\boldsymbol e_z\)
the frame's angular-velocity vector. The rotating-frame launch velocity
subtracts \(\boldsymbol n_p\times\boldsymbol r_0\).

Because the fixed radial distance calculation assigns the entire entry speed of
equation~\eqref{eq:legacy-5} to normal gas-relative motion, the orbital
first-pass result need not equal the fixed radial distance result. We isolate the
contribution of returns by comparing the first connected residence in
the gas domain with the cumulative delivery of the same trajectories.
After each exit we record size, velocity, planet-centered energy, and
angular momentum; a negative instantaneous two-body energy alone does
not end a trajectory. Later gas entries follow from the orbit, and each
incident fragment is counted once.

\subsection{Disk capture and Saturn absorption}\label{disk-capture-and-saturn-absorption}

Let \(r_p\), \(r_a\), and \(i_{\rm orb}\) be the pericenter, apocenter,
and inclination of the remnant's osculating Kepler orbit. Because the gas boundaries use
cylindrical radius, define the conservative lower bound \(R_{\rm lb}\)
by

\begin{equation}
R\geq r|\cos i_{\rm orb}|\geq r_p|\cos i_{\rm orb}|\equiv R_{\rm lb}.
\label{eq:legacy-12d}
\end{equation}

We distinguish disk capture from Saturn absorption.\footnote{In its one-dimensional,
perpendicular disk-passage calculation, \citetalias{mosqueira2010}
(Section 3.1) defined local capture by \(v_{\rm final}=0\), where
\(v_{\rm final}\) is the final speed along the passage.}
A surviving parent is captured by the disk when its osculating orbit is
bound and satisfies

\begin{equation}
\begin{aligned}
R_{\rm lb}&\geq2R_S,\qquad r_a\leq100R_S,\\
r_a\sin i_{\rm orb}&<H(R_{\rm lb}).
\end{aligned}
\label{eq:legacy-12e}
\end{equation}

These conditions place the osculating Kepler orbit wholly within the gas
domain. The vertical condition is conservative because \(H\) increases
with cylindrical radius. Eccentric and retrograde bound orbits are
included. Pressure-supported gas determines drag and heating throughout
the integration. Appendix~\ref{a.11.-vertical-extent} tests a wider vertical domain with the
same containment prescription.

We credit the surviving parent's mass at the position where disk
capture is first reached, which defines its radial mass assignment.
Vapor and retained melt are accumulated at their actual removal
locations. Disk capture ends the delivery calculation for that parent;
its subsequent damping, migration, and growth belong to the evolution
of the captured population.

A remnant crossing the spherical inner boundary \(r=2R_S\) inward is
instead counted as Saturn absorption; its remaining mass is recorded
separately and is not placed in a disk annulus. Disk delivery therefore
consists of vapor, retained melt, and disk-captured parents.

Histories are continued to the disk-capture endpoint,
inner-boundary absorption, complete ablation, or outward crossing of one
Hill radius. The Hill radius, \(R_H\), is

\[
R_H=\left(\frac{GM_S}{3n_p^2}\right)^{1/3}\simeq1.1\times10^3R_S.
\]

Outward crossing of this boundary ends the modeled encounter.

\subsection{Incoming ensemble and passage probabilities}\label{incoming-ensemble-and-passage-probabilities}

To measure first crossings and subsequent returns for a common source,
we launch an incoming ensemble at the Hill boundary. Orbital
orientations are isotropic. Let \(b_\infty\) denote the asymptotic impact
parameter, and let \(R_{\max}=100R_S\), or about \(0.1R_H\), be the
largest nominal two-body pericenter. We sample \(b_\infty^2\) uniformly over
\(0\leq b_\infty^2\leq R_{\max}^2+2GM_SR_{\max}/v_\infty^2\): this is
the focused encounter cross section for nominal two-body pericenters
inside \(100R_S\). At the Hill surface, let \(v\) be the total inertial
speed and \(v_t,v_r\) its tangential and signed radial velocity
components. We use \(v^2=v_\infty^2+2GM_S/R_H\),
\(v_t=b_\infty v_\infty/R_H\), and the inward radial velocity is
\(v_r=-(v^2-v_t^2)^{1/2}\). Hill motion then determines whether and
where the projectile first intersects the gas domain. Encounters that
miss it remain in the probability denominator. Four independently
scrambled Sobol samples specify the impact parameter and three
orientation angles; sizes and materials are initially independent of
these coordinates. The explicit isotropic coordinate map is given in
Appendix~\ref{a.8.-orbital-integration-and-conditional-probability-checks}. This distribution defines the incoming population used here.

For this ensemble, \(\Pr\) denotes probability, the vertical bar
denotes conditioning, and integer \(k\) numbers gas passages. Events are
written in sans-serif type: \(\mathsf F\) is first gas entry,
\(\mathsf X_k\) exit after passage \(k\), \(\mathsf N_{k+1}\) the
subsequent return, and \(\mathsf K_k\) a surviving parent reaching disk
capture or Saturn absorption during passage \(k\) or the ballistic
segment that follows it; \(\mathsf K\) denotes reaching either endpoint
on any passage. A remnant is counted in \(\mathsf K\) when it reaches
one of these endpoints; becoming gravitationally bound alone does not
end its tracked history. We measure
\(\Pr(\mathsf F)\), \(\Pr(\mathsf K_1\mid\mathsf F)\),
\(\Pr(\mathsf N_2\mid\mathsf X_1)\), \(\Pr(\mathsf K_2\mid\mathsf N_2)\),
and \(\Pr(\mathsf K\mid\mathsf N_2)\). Transitions after later exits use
the already processed survivors, with their changed sizes and
velocities. These probabilities refer to surviving parent solids;
complete ablation is recorded separately.

Object probabilities weight \(dN\propto s_0^{-p}ds_0\), whereas mass
efficiencies weight \(s_0^{3-p}ds_0\); a surviving-mass-weighted return
fraction also weights each exit by its remaining mass. Keeping these
denominators separate prevents first crossers, processed returners, and
delivered mass from being conflated.
Section~\ref{orbital-delivery-and-the-compositional-gradient} reports
the results; numerical, boundary, spatial, gas-rotation, and
vertical-extent checks are in
Appendices~\ref{a.8.-orbital-integration-and-conditional-probability-checks}--\ref{a.11.-vertical-extent}.

\section{From fragment efficiencies to a disk
composition}\label{from-fragment-efficiencies-to-a-disk-composition}

Sections~\ref{from-fragment-efficiencies-to-a-disk-composition}--\ref{sufficient-material-at-the-relevant-locations} use the fixed radial distance reference model. Section~\ref{orbital-delivery-and-the-compositional-gradient}
gives the separate orbital calculation, whose deposition radius can
differ from its entry radius.

Denote the component-normalized, mass-weighted delivery efficiency by
\(\overline D_j(R)\):

\begin{equation}
\overline D_j(R)=
\frac{\displaystyle\int_{s_{\min}}^{s_{\max}}
D_j(R,s_0)s_0^{4-p}\,d\ln s_0}
{\displaystyle\int_{s_{\min}}^{s_{\max}}s_0^{4-p}\,d\ln s_0}.
\label{eq:legacy-13}
\end{equation}

The same weighting gives the vapor, retained-melt, and confined-remnant
contributions. This is the component-normalized form of
\citetalias{mosqueira2010}'s size integral. Let \(\Sigma_{\rm pass}(R)\)
be incident fragment mass per unit disk area, \(\Sigma_i\) and
\(\Sigma_r\) the deposited ice and rock columns, and \(f_{i,\rm dep}\)
the deposited ice mass fraction. Then

\begin{equation}
\begin{aligned}
\Sigma_i&=f_0\Sigma_{\rm pass}\overline D_i,\qquad
\Sigma_r=(1-f_0)\Sigma_{\rm pass}\overline D_r,\\
f_{i,\rm dep}(R)&=\frac{f_0\overline D_i}
{f_0\overline D_i+(1-f_0)\overline D_r}.
\end{aligned}
\label{eq:legacy-14}
\end{equation}

Equation~\eqref{eq:legacy-14} gives both the enrichment relative to the
source and its variation with radius. For a nonzero deposited rock
column, define the ice-to-rock enrichment factor \(\mathcal F_{i/r}\) by

\begin{equation}
\mathcal F_{i/r}(R)\equiv
\frac{\Sigma_i(R)/\Sigma_r(R)}{f_0/(1-f_0)}
=\frac{\overline D_i(R)}{\overline D_r(R)}.
\label{eq:legacy-14a}
\end{equation}

Values of \(\mathcal F_{i/r}>1\) measure enrichment relative to the
source; variation with \(R\) measures the radial gradient. A common
incident normalization cancels from these ratios but remains in the
deposited mass.

Let \(\Sigma_\star\) be the common incident-column amplitude and
\(g(X)\) its dimensionless radial shape. For the incident column we
adopt \citetalias{mosqueira2010}'s focusing form,

\begin{equation}
\begin{gathered}
\Sigma_{\rm pass}(X)=\Sigma_\star g(X),\\
g(X)=1+\frac{2\Theta}{X},\qquad X=R/R_S.
\end{gathered}
\label{eq:legacy-15}
\end{equation}

The amplitude \(\Sigma_\star\) sets the amount of incoming material and
is the same for both components and all annuli. Let
\(X_1,X_2\) bound a radial interval, \(\Delta M_j\) be the component
mass deposited there, and \(f_j\) its incoming mass fraction, with
\(f_i=f_0\) and \(f_r=1-f_0\). Then:

\begin{equation}
\begin{gathered}
\Delta M_j(X_1,X_2)=\\
2\pi R_S^2\Sigma_\star f_j
\int_{X_1}^{X_2}Xg(X)\overline D_j(X)\,dX,\\
f_i=f_0,\qquad f_r=1-f_0.
\end{gathered}
\label{eq:legacy-16}
\end{equation}

Changing \(\Sigma_\star\) scales both component masses without changing
their ratio. With this amplitude held fixed, the mass in a local
interval does not depend on how the assumed profile is continued farther
out. The local mass-sufficiency test therefore uses the selected
interval directly.

\section{Fixed radial distance delivery and disk
composition}\label{calculated-delivery-and-compositional-gradients}

\subsection{Size-dependent delivery}\label{size-dependent-delivery}

\begin{figure*}[t!]
\begin{minipage}[t]{0.48\textwidth}
\vspace{0pt}
\centering
\includegraphics[width=1\linewidth,keepaspectratio,alt={Size-dependent local delivery at 20R\_S from the coupled crossing equations. The ordinate is deposited mass divided by initial fragment mass. Melt is credited under unit retention; confinement counts only the surviving parent. All curves use the same entry conditions.}]{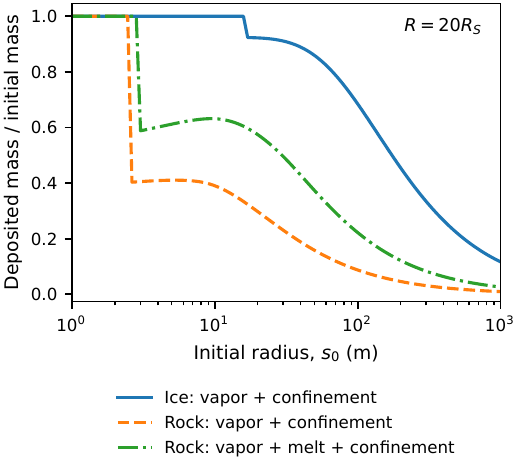}
\caption{Size-dependent local delivery at \(20R_S\) from the coupled
crossing equations. The ordinate is deposited mass divided by initial
fragment mass. Melt is credited under unit retention; confinement counts
only the surviving parent. All curves use the same entry
conditions.}\label{fig:size-titan}
\end{minipage}\hfill
\begin{minipage}[t]{0.48\textwidth}
\vspace{0pt}
\centering
\includegraphics[width=1\linewidth,keepaspectratio,alt={Local delivery at 60R\_S. The gap between the ice total and vapor-loss curves is confined-remnant mass. Ice sublimation supplies throughgoing fragments' contribution; rock delivery is negligible in both treatments.}]{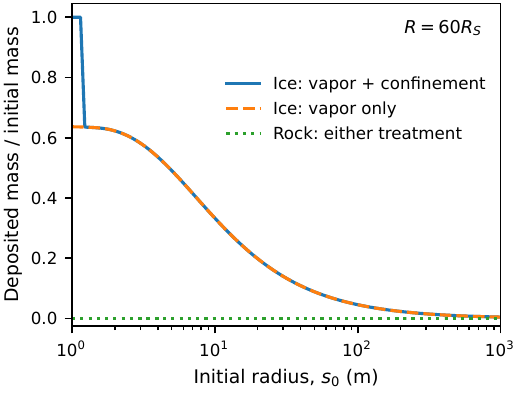}
\caption{Local delivery at \(60R_S\). The gap between the ice total and
vapor-loss curves is confined-remnant mass. Throughgoing ice fragments
contribute by sublimation; rock delivery is negligible in both
treatments.}\label{fig:size-iapetus}
\end{minipage}
\end{figure*}

At \(20R_S\) (Figure~\ref{fig:size-titan}), the largest initial radii delivering their full
mass locally are approximately 16 m for ice, 2.6 m for vapor-only rock,
and 2.9 m for rock with retained melt. Below these boundaries,
ablation and confinement account for the whole fragment. Larger bodies
emerge with surviving mass but deposit material during the crossing. The
sharp changes reflect the confinement definition: an emerging remnant
is not credited to that annulus. Melt shedding increases refractory
delivery by throughgoing bodies and changes their subsequent drag response.

At \(60R_S\) (Figure~\ref{fig:size-iapetus}), ice
fragments larger than approximately 1.2 m are not locally confined but
still deposit ice by sublimation. Rocky fragments of 1 m and larger are
neither confined nor appreciably ablated at the reference column. The
melt branch is inactive there.

\subsection{Size-integrated delivery
channels}\label{size-integrated-delivery-channels}

Table~\ref{tab:2} integrates over the reference \(p=3.5\) distribution. Each entry
is a fraction of the incoming mass of that component, before applying
the source fractions 0.30 for ice and 0.70 for rock.

\begin{table*}[t!]
\centering
\caption{Mass-weighted deposition channels at confinement or exit. Melt is
credited under unit local retention; confined mass is the surviving
parent only. Channels and totals are rounded independently from the
unrounded calculation. The refractory vapor proxy gives less than
\(10^{-9}\) delivery at \(60R_S\), treated as negligible.\label{tab:2}}
\small\setlength{\tabcolsep}{4pt}\renewcommand{\arraystretch}{1.15}
\begin{tabular}{@{}>{\raggedright\arraybackslash}p{(\linewidth - 10\tabcolsep)*\real{0.06500000}}>{\raggedright\arraybackslash}p{(\linewidth - 10\tabcolsep)*\real{0.33500000}}>{\raggedright\arraybackslash}p{(\linewidth - 10\tabcolsep)*\real{0.13000000}}>{\raggedright\arraybackslash}p{(\linewidth - 10\tabcolsep)*\real{0.14000000}}>{\raggedright\arraybackslash}p{(\linewidth - 10\tabcolsep)*\real{0.16000000}}>{\raggedright\arraybackslash}p{(\linewidth - 10\tabcolsep)*\real{0.17000000}}@{}}
\toprule
\(R/R_S\) & Material treatment & Vapor & Shed melt & Confined remnant & Total delivered\\
\midrule
20 & Ice & 0.46 & 0 & 0.0075 & 0.47\\
20 & Rock, vapor only & 0.094 & 0 & 0.012 & 0.11\\
20 & Rock, retained melt & 0.033 & 0.16 & 0.0099 & 0.20\\
60 & Ice & 0.075 & 0 & 0.0010 & 0.076\\
60 & Rock, either treatment & Negligible & 0 & 0 & Negligible\\
\bottomrule
\end{tabular}
\end{table*}

At \(20R_S\), retained melt shedding increases total rock delivery from
11\% to 20\% of the incoming rock. Vapor loss decreases because shedding
holds the surface at a lower temperature and changes the trajectory. Of
the deposited rock in the shedding case, 79\% is retained melt, 17\%
vapor, and 4.9\% confined parent remnants. Channel percentages are rounded
independently and need not sum to exactly 100\%. The increase therefore comes
mainly from shed melt assumed to be retained, not from confinement of
more intact parents.

At \(60R_S\), the deposit receives about 7.6\% of the incoming ice mass;
99\% of that deposit comes from sublimation and 1.3\% from confined
parent remnants. Rock delivery is negligible. Differential thermal
ablation therefore drives the modeled delivery gradient, although drag remains
coupled to every trajectory.

\subsection{Ice enrichment and radial composition of the
deposit}\label{bulk-enrichment-and-the-radial-compositional-gradient}

\begin{figure*}[t!]
\begin{minipage}[t]{0.48\textwidth}
\vspace{0pt}
\centering
\includegraphics[width=1\linewidth,keepaspectratio,alt={Deposited component columns for \textbackslash Sigma\_\{\textbackslash rm pass\}=\textbackslash Sigma\_\textbackslash star(1+200/X), including source mass fractions. The ice column is common to both treatments; melt is retained under equation~\eqref{eq:legacy-12a}. Values below 10\^{}\{-5\} are omitted from the plot, not replaced by a floor in the calculation.}]{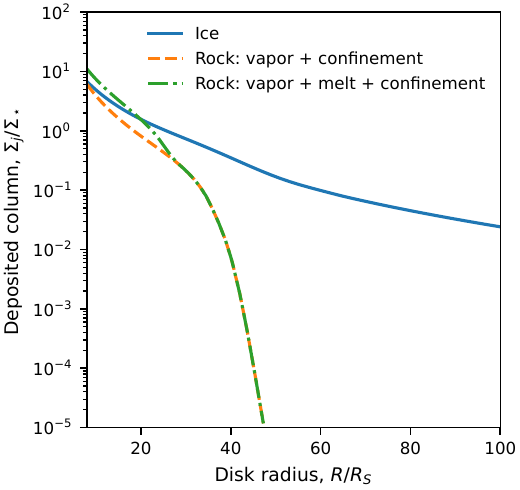}
\caption{Deposited component columns for
\(\Sigma_{\rm pass}=\Sigma_\star(1+200/X)\), including source mass
fractions. The ice column is common to both treatments; melt is retained
under equation~\eqref{eq:legacy-12a}. Values below \(10^{-5}\) are omitted from the
plot, not replaced by a floor in the calculation.}\label{fig:columns}
\end{minipage}\hfill
\begin{minipage}[t]{0.48\textwidth}
\vspace{0pt}
\centering
\includegraphics[width=1\linewidth,keepaspectratio,alt={Radial deposited ice fraction with pressure-aware Titan and compact, zero-porosity Iapetus comparisons. The deposit is enriched relative to the 0.30 source line in the inner disk and becomes nearly pure ice farther out. Titan's marker is the midpoint of the 40 to 50 percent water-rich range of Sohl et al. 2003; Iapetus uses the compact reference calculation of Appendix~\ref{a.7.-size-data-and-the-compact-reference-calculations}.}]{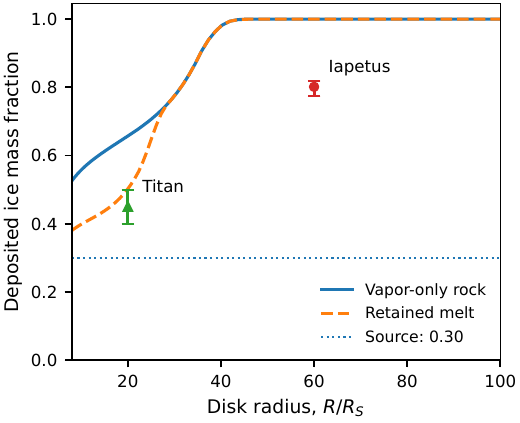}
\caption{Radial deposited ice fraction in the fixed radial distance reference.
The dotted line is the 0.30 source fraction. The retained-melt curve
assumes local retention of shed melt. Markers show the pressure-aware
Titan composition (midpoint of the 40--50\% water-rich range of
\citealp{sohl2003}) and the compact, zero-porosity Iapetus reference
(Appendix~\ref{a.7.-size-data-and-the-compact-reference-calculations}) at their present orbits, for context. Their intervals are
those of the interior models.}\label{fig:gradient}
\end{minipage}
\end{figure*}

The fixed radial distance calculation produces an ice-enriched deposit, with
the strongest compositional selection in the outer region. Figure~\ref{fig:columns} gives the deposited component columns:
the ice fraction rises outward because
refractory delivery declines more rapidly than ice delivery. Figure~\ref{fig:gradient}
shows that the deposit is already enriched relative to the source in the
inner disk, before the further increase outward:

\begin{equation}
\begin{gathered}
f_{i,\rm dep}(20R_S)=
\begin{cases}
0.66,&\text{vapor-only rock},\\
0.50,&\text{retained melt},
\end{cases}\\
f_{i,\rm dep}(60R_S)\simeq1.
\end{gathered}
\label{eq:legacy-17}
\end{equation}

The corresponding inner ice-to-rock enrichment factors are 4.5 and 2.4
(equation~\eqref{eq:legacy-14a}). Even when retained melt strengthens refractory
delivery, the inner deposit remains enriched relative to the 30/70
ice--rock source. Enrichment relative to the source is therefore not
confined to the outer disk. At \(40R_S\), both treatments give
\(f_{i,\rm dep}=0.98\), and melt shedding no longer matters.

For context, Figure~\ref{fig:gradient} also shows the composition references of Figure~\ref{fig:observational-context}.
The pressure-aware Titan models of \citet{sohl2003} give 40--50\% water-rich material by
mass (midpoint 45\%). For Iapetus, we use the compact reference calculation of
Appendix~\ref{a.7.-size-data-and-the-compact-reference-calculations}, about 80\% water by mass. These references are independent
of the fragment calculation and are not fitted to it (Appendix~\ref{a.4.-composition-references-and-the-saturn-disk-overlays}).

The radial variation follows from the different thermal responses of
ice and rock. Rock vaporization and retained melt contribute in the
inner disk, while ice sublimation continues farther out after
significant refractory delivery
has ceased. The near-pure-ice outer limit is an end-member outcome. The
next two subsections show how the gas column and the size weighting
change the accompanying rock and the delivered mass.

\subsection{Confinement of rocky remnants at higher outer-disk
columns}\label{capture-of-rocky-remnants-at-higher-outer-disk-columns}

\begin{figure*}[t!]
\begin{minipage}[t]{0.48\textwidth}
\vspace{0pt}
\centering
\includegraphics[width=1\linewidth,keepaspectratio,alt={Outer-disk ice fraction when only the gas column at 60R\_S varies. The two rock treatments coincide. Rocky-remnant confinement supplies most of the accompanying rock at higher columns. The shaded range and dashed line show the zero-porosity Iapetus reference of Figures~\ref{fig:observational-context} and \ref{fig:gradient} for context.}]{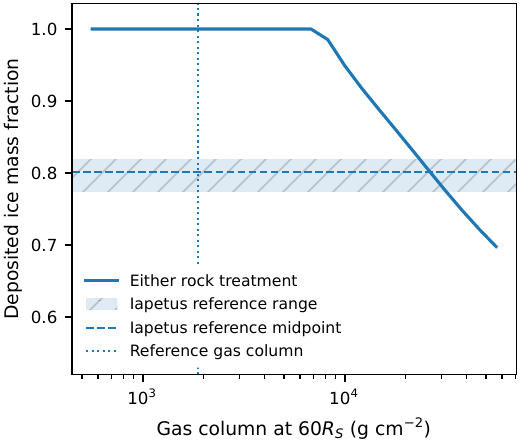}
\caption{Outer-disk ice fraction when only the gas column at \(60R_S\)
varies. The two rock treatments coincide. Rocky-remnant confinement supplies
most of the accompanying rock at higher columns. The shaded range and
dashed line show the zero-porosity Iapetus reference of Figures~\ref{fig:observational-context} and \ref{fig:gradient}
for context.}\label{fig:column-comparison}
\end{minipage}\hfill
\begin{minipage}[t]{0.48\textwidth}
\vspace{0pt}
\centering
\includegraphics[width=1\linewidth,keepaspectratio,alt={Modeled radial deposited ice fractions for p=3, 3.5, and 4 with retained melt. Each source has 1--1000 m radii and aggregate ice fraction 0.30. The guides mark the present orbits of Titan and Iapetus. Outer ice enrichment persists while the inner composition and delivered mass change.}]{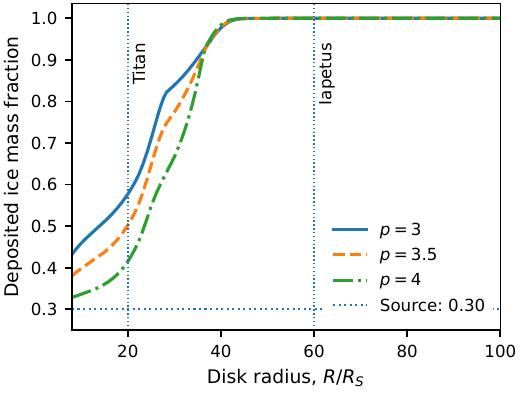}
\caption{Modeled radial deposited ice fractions for \(p=3\), 3.5, and 4
with retained melt. Each source contains fragments with radii of 1--1000 m
and an aggregate ice fraction of 0.30. The guides mark the present orbits of Titan and Iapetus. Outer ice enrichment
persists while the inner composition and delivered mass
change.}\label{fig:powerlaw}
\end{minipage}
\end{figure*}

Figure~\ref{fig:column-comparison} varies only the gas column at \(60R_S\), holding the
temperature, scale height, entry speed, size distribution, and source
ratio fixed. At ten times the reference column,
\(\Sigma_g=1.9\times10^4\ \mathrm{g\,cm^{-2}}\),

\begin{equation}
\overline D_i=0.28,\qquad
\overline D_r=0.020,\qquad
f_{i,\rm dep}=0.85.
\label{eq:legacy-19}
\end{equation}

Confined remnants supply 0.018 of incoming rock mass, about 90\% of the
deposited rock; the remainder is vapor. At thirty times the reference
column, \(f_{i,\rm dep}=0.70\). The two rock treatments coincide:
confinement admits additional rock without activating melt shedding.
Higher outer columns therefore moderate the enrichment through
confinement of rocky remnants, a different channel from the refractory
ablation that dominates the inner disk.

\subsection{Dependence on the size-distribution index}\label{dependence-on-the-power-law-slope}

Reweighting the same fragment calculations with \(p=3\), 3.5, and 4
gives \(20R_S\) ice fractions of 0.58, 0.50, and 0.42 with retained
melt (Figure~\ref{fig:powerlaw}). The vapor-only values are 0.76, 0.66, and 0.51. All
exceed the 0.30 source fraction. A steeper distribution assigns more mass
to small fragments, which are delivered efficiently, and this raises
refractory delivery in the inner disk.

All three indices give nearly pure-ice deposition at \(60R_S\), but the
fractions of incoming ice delivered there are 0.025, 0.076, and 0.25.
Persistent enrichment therefore does not imply a fixed inventory. The
mass calculation below retains the same component weighting.

\section{Sufficient material at the relevant
locations}\label{sufficient-material-at-the-relevant-locations}

An ice-rich deposit must also contain sufficient mass. Equation~\eqref{eq:legacy-16}
supplies a local test using the reference \(p=3.5\) population and one
incident-column amplitude,

\begin{equation}
\Sigma_\star\simeq5.6\times10^2\ \mathrm{g\,cm^{-2}}.
\label{eq:legacy-20}
\end{equation}

\begin{figure*}[t!]
\begin{minipage}[t]{0.48\textwidth}
\vspace{0pt}
\centering
\includegraphics[width=1\linewidth,keepaspectratio,alt={Mass deposited between 10R\_S and the plotted radius. At 32R\_S, the two rock treatments supply the Titan-scale inventories in Table~\ref{tab:3}. The horizontal line marks one Titan mass.}]{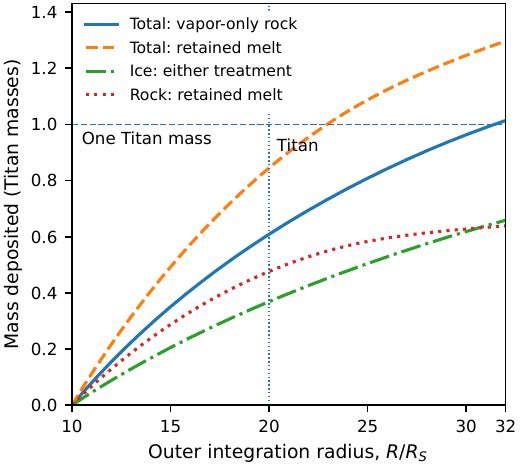}
\caption{Mass deposited between \(10R_S\) and the plotted radius. At
\(32R_S\), the two rock treatments supply the Titan-scale inventories in
Table~\ref{tab:3}. The horizontal line marks one Titan
mass.}\label{fig:local-titan}
\end{minipage}\hfill
\begin{minipage}[t]{0.48\textwidth}
\vspace{0pt}
\centering
\includegraphics[width=1\linewidth,keepaspectratio,alt={Mass deposited from 60R\_S outward, in Iapetus units, with the same incident amplitude as Figure~\ref{fig:local-titan}. The treatments coincide. The 60--63R\_S interval contains about 1.2M\_I of ice-rich material without extrapolation to the tidal boundary.}]{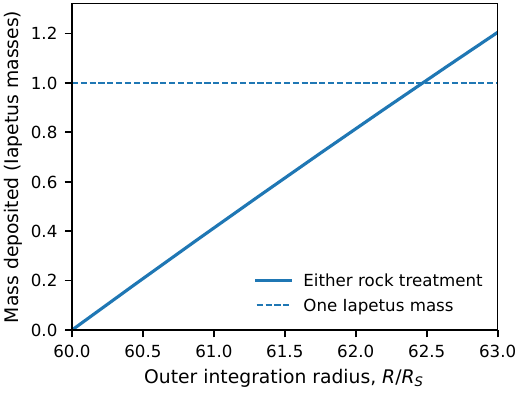}
\caption{Mass deposited from \(60R_S\) outward, in Iapetus units, with
the same incident amplitude as Figure~\ref{fig:local-titan}. The treatments coincide. The
\(60\)--\(63R_S\) interval contains about \(1.2M_I\) of ice-rich
material without extrapolation to the tidal
boundary.}\label{fig:local-iapetus}
\end{minipage}
\end{figure*}

Appendix~\ref{a.5.-independent-checks-and-mass-integration} gives the equivalent source normalization. Table~\ref{tab:3} and
Figures~\ref{fig:local-titan}--\ref{fig:local-iapetus} use it in both regions without changing disk, fragment, or
thermal parameters. The comparison masses of Titan and Iapetus, \(M_T\)
and \(M_I\), are

\begin{equation}
M_T\simeq1.3\times10^{23}\ \mathrm{kg},\qquad
M_I\simeq1.8\times10^{21}\ \mathrm{kg},
\label{eq:legacy-21}
\end{equation}

rounded here from the JPL SAT441 gravitational parameters
\citep{jpl2026a,jpl2026b}, independently of the density-to-composition
conversion. Appendix~\ref{a.5.-independent-checks-and-mass-integration} records the unrounded mass scales used in the
inventory calculation.

\begin{table*}[t!]
\centering
\caption{Available local mass and its mass-weighted ice fraction at \(p=3.5\) and
the common amplitude in equation~\eqref{eq:legacy-20}. The two rock treatments are
alternatives, not contributions to be summed. The vapor-only inner
inventory rounds to one Titan mass; the unrounded value lies slightly
above it.\label{tab:3}}
\small\setlength{\tabcolsep}{4pt}\renewcommand{\arraystretch}{1.15}
\begin{tabular}{@{}>{\raggedright\arraybackslash}p{(\linewidth - 6\tabcolsep)*\real{0.19000000}}>{\raggedright\arraybackslash}p{(\linewidth - 6\tabcolsep)*\real{0.35000000}}>{\raggedright\arraybackslash}p{(\linewidth - 6\tabcolsep)*\real{0.25000000}}>{\raggedright\arraybackslash}p{(\linewidth - 6\tabcolsep)*\real{0.21000000}}@{}}
\toprule
Radial interval & Rock treatment & Available mass & Ice mass fraction\\
\midrule
\(10\)--\(32R_S\) & Vapor-only & \(1.0M_T\) & 0.65\\
\(10\)--\(32R_S\) & Retained melt & \(1.3M_T\) & 0.51\\
\(60\)--\(63R_S\) & Either treatment & \(1.2M_I\) & \(\simeq1\)\\
\bottomrule
\end{tabular}
\end{table*}

Because mass in a local interval is fixed by \(\Sigma_\star\),
conditions near the tidal boundary need not be specified; extending
the assumed profiles outward changes only the extrapolated total.

\section{Orbital delivery and radial disk
composition}\label{orbital-delivery-and-the-compositional-gradient}

\subsection{First encounters, returns, and cumulative
capture}\label{first-encounters-returns-and-cumulative-capture}

The orbital extension distinguishes incoming encounters from already
processed survivors. Of 512 Hill-boundary trajectories, 422 first enter
the gas domain, giving \(\Pr(\mathsf F)=0.82\) for the specified focused,
isotropic encounter ensemble. Each successful geometry is followed for
64 logarithmic size bins and all three material treatments: 81,024
coupled size-material histories. Four independent source scrambles
provide a sampling check. A remnant that leaves the disk is propagated
with its actual exit state; subsequent entry and capture are outcomes of
that orbit. Four additional Hill-boundary trajectories cross \(2R_S\)
without first encountering gas; they count as direct Saturn absorption
and are outside the conditional first-crosser statistics. Table~\ref{tab:4}
reports the conditional passage and endpoint probabilities.

\begin{table*}[t!]
\centering
\caption{Object-count probabilities for \(p=3.5\), pressure-supported gas rotation. The
four probability columns are \(\Pr(\mathsf K_1\mid\mathsf F)\),
\(\Pr(\mathsf N_2\mid\mathsf X_1)\), \(\Pr(\mathsf K_2\mid\mathsf N_2)\),
and \(\mathsf K\) on any passage from the second onward conditional on
\(\mathsf N_2\), respectively. Their denominators differ.
Histories are followed to a physical endpoint, with an absorbing
boundary at \(2R_S\) and an outer exit at one Hill radius. \(\mathsf K\) counts disk capture or Saturn absorption; complete ablation is recorded separately. Rock treatments are alternatives. Small fragments
dominate these object counts; the mass-weighted results below use the
same histories with different weights.\label{tab:4}}
\small\setlength{\tabcolsep}{4pt}\renewcommand{\arraystretch}{1.15}
\begin{tabular}{@{}>{\raggedright\arraybackslash}p{(\linewidth - 8\tabcolsep)*\real{0.24000000}}>{\raggedright\arraybackslash}p{(\linewidth - 8\tabcolsep)*\real{0.19000000}}>{\raggedright\arraybackslash}p{(\linewidth - 8\tabcolsep)*\real{0.19000000}}>{\raggedright\arraybackslash}p{(\linewidth - 8\tabcolsep)*\real{0.19000000}}>{\raggedright\arraybackslash}p{(\linewidth - 8\tabcolsep)*\real{0.19000000}}@{}}
\toprule
Material treatment & \(\mathsf K\) after one passage & Return after first exit & \(\mathsf K\) on next return & \(\mathsf K\) during or after return\\
\midrule
Ice & 0.54 & 0.82 & 0.75 & 0.98\\
Rock: vapor-only & 0.25 & 0.70 & 0.50 & 0.97\\
Rock: retained melt & 0.24 & 0.70 & 0.48 & 0.95\\
\bottomrule
\end{tabular}
\end{table*}

For vapor-only rock, 70\% of the first-exiting objects return, but they
carry only 26\% of the surviving first-exit mass. Among returning
objects, 50\% reach disk capture or Saturn absorption on the next
passage and 97\% eventually reach one of these endpoints. For ice, 54\%
of first crossers are disk-captured or absorbed by Saturn during their
first passage; among returning icy objects, 75\% reach one of these
endpoints on the next passage and 98\% eventually do so, while 1.1\%
fully ablate. Complete ablation supplies mass to the disk but is counted
separately from \(\mathsf K\).

Across the four source scrambles, the first-entry probability ranges
from 0.79 to 0.85, the rock return probability from 0.68 to 0.72, and
\(\Pr(\mathsf K\mid\mathsf N_2)\) from 0.95 to 0.98; these ranges
reflect finite encounter sampling. Returns select a different population
from first crossers, so conditional populations are tracked separately
after the first, second, third, fourth, and subsequent exits. Applying a
fixed return probability anew to the original size distribution would
lose this selection and count already removed material again.

\begin{figure*}[t!]
\centering
\includegraphics[width=1\linewidth,keepaspectratio,alt={Orbital continuation and cumulative delivery in the pressure-supported disk. (a) A 3 m rocky fragment in the vertical-entry example starts at 60R\_S, returns five times, and reaches disk capture near 26R\_S; colors distinguish the six gas visits and their subsequent ballistic segments. The shaded band is the modeled gas domain; the dot and star mark entry and capture. (b) Conditional object probabilities for the incoming Hill ensemble; K denotes disk capture or Saturn absorption, with denominators defined in Table~\ref{tab:4}. (c) First-pass and additional mass delivery, normalized separately to each component's first-crosser mass. (d) Cumulative deposition at the actual removal or disk-capture radius, normalized to the common encounter mass and weighted by the 30/70 ice-rock source ratio. Dashed curves include first passages only; solid curves include returns. The shaded interval is 60--63R\_S. Panels (b)--(d) use p=3.5 and the independently sampled incoming ensemble, rather than the individual orbit in (a).}]{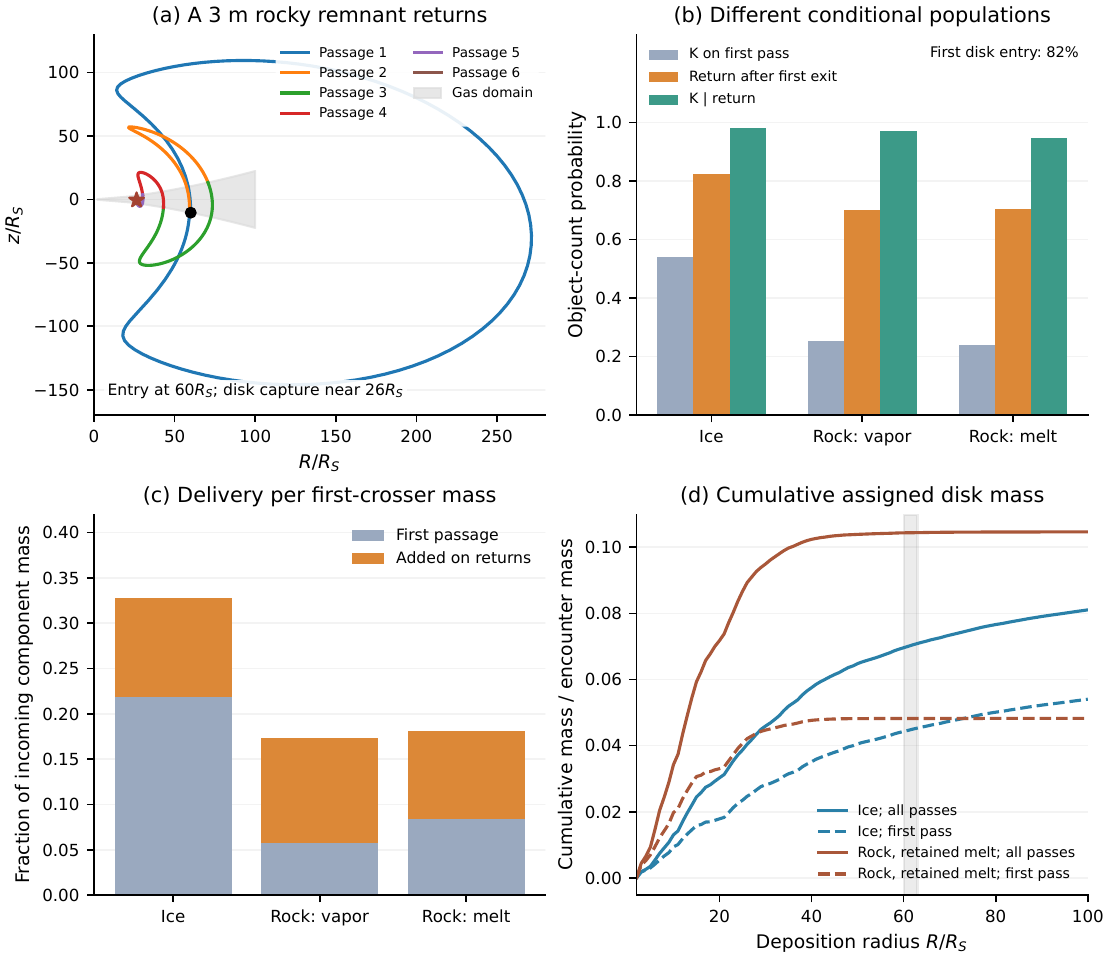}
\caption{Orbital continuation and cumulative delivery in the
pressure-supported disk. (a) A 3 m rocky fragment in the vertical-entry
example starts at \(60R_S\), returns five times, and reaches disk capture near
\(26R_S\); colors distinguish the six gas visits and
their subsequent ballistic segments. The shaded band is the modeled gas
domain; the dot and star mark entry and capture. (b) Conditional object probabilities for the incoming Hill ensemble;
\(\mathsf K\) denotes disk capture or Saturn absorption, with denominators
defined in Table~\ref{tab:4}. (c) First-pass and additional mass delivery,
normalized separately to each component's first-crosser mass. (d)
Cumulative deposition at the actual removal or disk-capture radius,
normalized to the common encounter mass and weighted by the 30/70
ice--rock source ratio. Dashed curves include first passages only; solid
curves include returns. The shaded interval is \(60\)--\(63R_S\). Panels
(b)--(d) use \(p=3.5\) and the independently sampled incoming ensemble,
rather than the individual orbit in (a).}\label{fig:multipass}
\end{figure*}

Normalized to the component mass that first crosses the disk, ice
delivery rises from 0.22 to 0.33 when later passages are included.
Vapor-only rock rises from 0.058 to 0.17, and rock with retained melt
from 0.084 to 0.18. The disk-captured parent contributions are
0.025, 0.052, and 0.037, respectively. Vapor contributes 0.30 for ice
and 0.12 for vapor-only rock. With retained melt, refractory vapor and
melt contribute 0.015 and 0.13, respectively. In addition,
surviving fractions of 0.0028, 0.0060, and 0.0040 cross \(2R_S\)
and count as Saturn absorption without contributing to disk deposition.
These channel fractions have the same first-crosser mass normalization
as the totals and are rounded independently. The first-pass and
cumulative values use the same trajectories, isolating later processing
from changes in encounter geometry. Figure~\ref{fig:multipass} shows the probability and
mass distinctions together.

The probabilities of \(\mathsf K\) include both surviving-parent
endpoints, while Figures~\ref{fig:multipass}--\ref{fig:orbital-composition} credit only disk-deposited material to
radial annuli. Pressure-supported gas drives inward transport: remnants
may reach disk capture at smaller radii or cross \(2R_S\) into Saturn's absorbing
inner boundary. The source-sampling bands show the variation across the four incoming
ensembles at fixed disk structure and radial assignment.

Individual controlled entries illustrate radial transport before disk capture.
A 3 m rock fragment entering vertically at \(60R_S\) reaches disk capture after
six gas passages near \(26R_S\), with almost all its mass intact. A 1 m
rock fragment entering there reaches disk capture after three passages near
\(30R_S\). A 10 m ice fragment supplies essentially all its initial
mass through ablation and disk capture, leaving a parent fraction of 0.18 at
\(28R_S\). An emerging remnant therefore cannot be assigned to its
first-crossing annulus. The Keplerian-gas control in Appendix~\ref{a.10.-gas-rotation-and-physical-boundary-sensitivities}
isolates the effect of pressure support using the same capture criterion.

Multiple passages thus change both the amount of refractory delivery and
where it is deposited. The nearly pure-ice outer deposit of the
single-crossing reference does not mean that rock entering the outer
disk escapes: surviving rock can return and be processed farther inward,
including absorption by Saturn at the inner boundary. The orbital
kernels therefore cannot be added to the fixed radial distance inventories of
Table~\ref{tab:3} as a uniform correction; instead, a common source
distribution must be propagated through the radial delivery kernel, as
in Figure~\ref{fig:multipass}(d).

\subsection{Ice fraction at the deposition
radius}\label{ice-fraction-at-the-deposition-radius}

The orbital composition profile follows from the mass delivered into
each annulus, including material arriving there after earlier encounters
elsewhere. Let \(K_{i,a}\) and \(K_{r,a}\) be the ice and rock delivery
fractions in annulus \(a\), normalized separately to each component's
incoming encounter mass and summed over vapor, retained melt where
applicable, and remnants at the disk-capture endpoint; inner-boundary absorption is
excluded. For source ice fraction \(f_0\),
denote the annular deposited ice fraction by \(f_{i,a}\):

\begin{equation}
f_{i,a}=\frac{f_0 K_{i,a}}{f_0 K_{i,a}+(1-f_0)K_{r,a}}.\label{eq:legacy-22}
\end{equation}

For any illustrative interval \(A\), let \(K_{j,A}\) be the sum of the
component kernels over its annuli and \(\epsilon_A\) its total deposited
fraction per unit incoming encounter mass:

\begin{equation}
K_{j,A}=\sum_{a\in A}K_{j,a},\qquad
\epsilon_A=f_0K_{i,A}+(1-f_0)K_{r,A}.
\label{eq:legacy-23}
\end{equation}

Let \(M_{\rm enc}\) denote the total incident encounter mass, including
encounters that miss the gas. Denote the deposited ice, rock, and total
masses in the interval by \(M_{i,A}\), \(M_{r,A}\), and
\(M_{{\rm dep},A}\), respectively. Then

\begin{equation}
\begin{aligned}
M_{i,A}&=M_{\rm enc}f_0K_{i,A},\\
M_{r,A}&=M_{\rm enc}(1-f_0)K_{r,A},\\
M_{{\rm dep},A}&=M_{\rm enc}\epsilon_A.
\end{aligned}
\label{eq:legacy-24}
\end{equation}

These relations give the mass delivered per unit incoming mass. Let \(M_{\rm pass}\)
be the distinct incident fragment mass over \(8\leq X\leq100\) in the
fixed radial distance calculation (Appendix~\ref{a.5.-independent-checks-and-mass-integration}). It normalizes a different
disk-crossing population; its numerical value cannot be assigned to
\(M_{\rm enc}\) without an additional source convention.

\begin{figure*}[t!]
\centering
\includegraphics[width=1\linewidth,keepaspectratio,alt={Ice fraction of the radially assigned disk inventory for a 30/70 ice-rock source and pressure-supported gas rotation. Columns show the two alternative rock treatments. Upper panels compare deposition on first gas passages alone (dashed) with all passages (solid) for p=3.5; shading spans the four independent encounter-scramble estimates for all passages. Lower panels reweight the same orbital histories with p=3, 3.5, and 4. Steps are ratios of component masses within a 2--8R\_S bin followed by 8R\_S annuli, with a 96--100R\_S final bin. The dotted horizontal line marks the incoming ice fraction. Vertical guides mark the present orbits of Titan and Iapetus. Disk delivery excludes Saturn absorption.}]{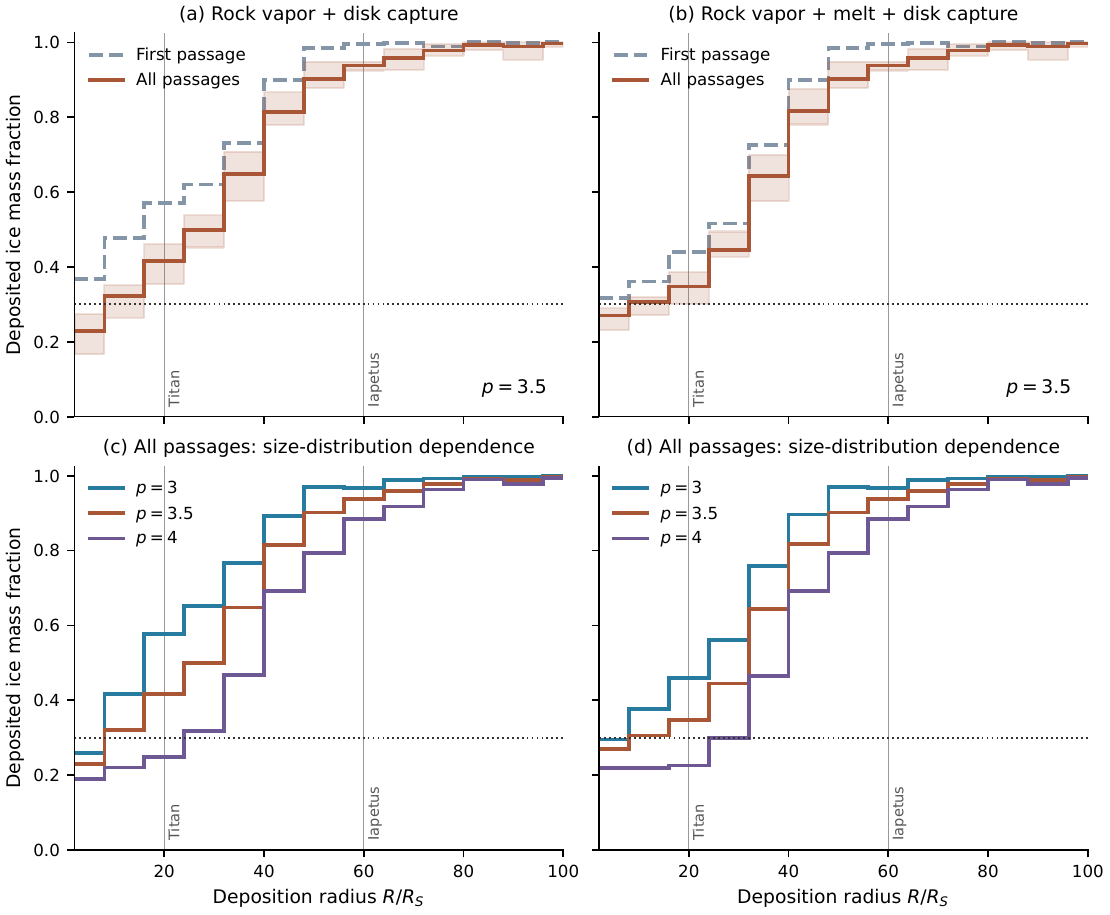}
\caption{Ice fraction of the radially assigned disk inventory for a 30/70
ice--rock source and pressure-supported gas rotation. Columns show the two
alternative rock treatments. Upper panels compare deposition on first
gas passages alone (dashed) with all passages (solid) for \(p=3.5\);
shading spans the range of the four independent encounter-scramble
estimates for all passages. Lower panels reweight the same orbital
histories with \(p=3\), 3.5, and 4. Steps are ratios of component masses
within an inner \(2\)--\(8R_S\) bin followed by \(8R_S\) annuli, with a \(96\)--\(100R_S\) final bin. The dotted
horizontal line marks the incoming ice fraction, and vertical guides mark
the present orbits of Titan and Iapetus. Disk delivery excludes
Saturn absorption.}\label{fig:orbital-composition}
\end{figure*}

The common annular area cancels in equation~\eqref{eq:legacy-22}. The first-pass profile
uses only deposition during the first gas visit of these same
trajectories; the all-pass profile adds subsequent visits without
resampling the source. These are local annular ratios, not ratios of
cumulative masses interior to \(R\), and not averages of individual
fragment compositions. Figure~\ref{fig:orbital-composition} uses an inner \(2\)--\(8R_S\) bin
and then \(8R_S\) annuli, with the final
interval truncated at \(100R_S\), to display the sampled spatial
structure. The calculation also evaluates the underlying \(1R_S\)
component kernels and \(4R_S\) and \(16R_S\) alternatives.

Across the full modeled domain, the all-pass deposited ice fraction is
0.45 for vapor-only rock and 0.44 with retained melt. Both exceed the
0.30 source fraction. Within \(10\)--\(32R_S\), the total deposition
efficiencies per unit encounter mass are 0.086 and 0.098, respectively;
in \(56\)--\(64R_S\) they are 0.0039 for either rock treatment.
Equation~\eqref{eq:legacy-24} converts these delivery efficiencies to masses for any explicitly
chosen \(M_{\rm enc}\), without equating it to the fixed radial distance normalization.

For \(p=3.5\), the \(16\)--\(24R_S\) annulus has first-pass ice
fractions of 0.57 for vapor-only rock and 0.44 with retained
melt. Including returns gives 0.42 and 0.35, respectively. In
\(24\)--\(32R_S\), the all-pass values are 0.50 and 0.44.
Over the broader \(10\)--\(32R_S\) reservoir they are 0.41 and
0.36. The orbital and fixed radial distance calculations differ in geometry
and normalization, so the effect of orbital continuation is measured
by the matched first-pass/all-pass comparison in Figure~\ref{fig:orbital-composition}, not by
comparison with the fixed radial distance values of 0.66 and 0.50.

The first-pass deposit is nearly pure ice in both \(56\)--\(64R_S\)
and \(60\)--\(63R_S\). After returns, their ice fractions are
0.94 and 0.96, respectively, for either rock treatment to the
quoted precision. Returned remnants supply rock to the outer disk,
which remains icier than both the source and the inner deposit.
The all-pass profiles preserve whole-disk enrichment and the outward
increase over the radial scale separating Titan and Iapetus. Reweighting
the same histories at \(p=3\) and 4 changes the level and radial shape
of the profile (lower panels of Figure~\ref{fig:orbital-composition}),
but the outward increase persists for all three size-distribution
indices.

The modeled delivery gradient does not imply enrichment above the source in
every annulus. For \(p=4\), the \(16\)--\(24R_S\) ice fractions are
0.25 and 0.23 for vapor-only rock and retained melt, respectively;
over \(10\)--\(32R_S\), they are 0.26 and 0.25. Even for
\(p=3.5\), the innermost \(2\)--\(8R_S\) bin has fractions of
0.23 and 0.27. The fixed radial distance reference and the fiducial orbital
population both enrich the Titan-region supply, but that local result
does not extend to every size-distribution index or radius.

\subsection{A corresponding Jupiter calculation}\label{jupiter-comparison}

We apply the same orbital calculation to Jupiter, retaining the
30/70 ice--rock source, 1--1000 m fragments, three size-distribution
indices, and both rock treatments. Let \(R_J\) and \(M_J\) denote
Jupiter's radius and mass, and define \(X_J=R/R_J\). We adopt
\citetalias{mosqueira2010}'s Jovian gas-column branches and temperature:

\begin{equation}
\begin{aligned}
\Sigma_{\rm in}&=2\times10^4\frac{15}{X_J}
\quad [\mathrm{g\,cm^{-2}}],\\
\Sigma_{\rm out}&=5\times10^3\left(\frac{25}{X_J}\right)^2
\quad [\mathrm{g\,cm^{-2}}],\\
T_g&=\max(3600/X_J,130)\ \mathrm{K}.
\end{aligned}
\label{eq:jupiter-disk}
\end{equation}

The branches join between \(15R_J\) and \(25R_J\) using the same
smooth interpolation as for Saturn, with
\(\xi=(X_J-15)/10\) clipped to \([0,1]\) in
equation~\eqref{eq:legacy-A1}. The Gaussian vertical structure and
pressure-supported rotation follow equations~\eqref{eq:legacy-3}
and~\eqref{eq:legacy-12c}, replacing \(M_S\) by \(M_J\).
The modeled gas domain is \(2\)--\(100R_J\) and \(|z|\leq H(R)\).
The outer boundary is again about \(0.1R_H\), with \(R_H\) now
Jupiter's Hill radius.
Disk capture uses the same bound, disk-contained criterion;
absorption at the spherical \(2R_J\) boundary is recorded separately.

The same 512 scrambled Sobol points specify impact parameters and
orientations at Jupiter's Hill sphere, with \(\Theta=100\) and nominal
two-body pericenters from 0 to \(100R_J\). We choose the same focusing
parameter as for Saturn to compare the planets at a common source
parameter; \citetalias{mosqueira2010} used \(\Theta=25\) for Jupiter
and 100 for Saturn. Appendix~\ref{a.jupiter-focusing} gives a matched
Jupiter calculation at \(\Theta=25\). Each first crosser receives
the same 64 size bins and three material treatments. The thermal laws,
orbital tolerances, and radial deposition sampling match those of the Saturn ensemble. The ballistic
first-entry search uses 32 subdivisions per step to resolve grazing
entries at the outer cylinder.
Appendix~\ref{a.jupiter-checks} gives the planetary constants
and numerical checks. The first-entry probability is 0.81. For
\(p=3.5\), the probability of returning after a first exit is 0.75
for ice and 0.55 or 0.54 for the two rock treatments; subsequent disk
capture or planetary absorption occurs for 0.96, 0.95, and 0.84 of
these returning objects, respectively.

\begin{figure*}[p]
\centering
\includegraphics[width=0.97\linewidth,alt={Four panels show the assigned ice mass fraction versus distance from Jupiter for a 30/70 ice-rock source. Left and right columns use vapor-only rock and retained rock melt, including captured remnants. Upper panels compare first passages with all passages for p=3.5; shading spans four source scrambles. Lower panels compare p=3, 3.5, and 4 using all passages. Ice fractions generally increase outward. Horizontal dotted lines mark the source fraction; vertical guides mark the Galilean satellite orbits. Delivered water includes vapor available as ice after condensation.}]{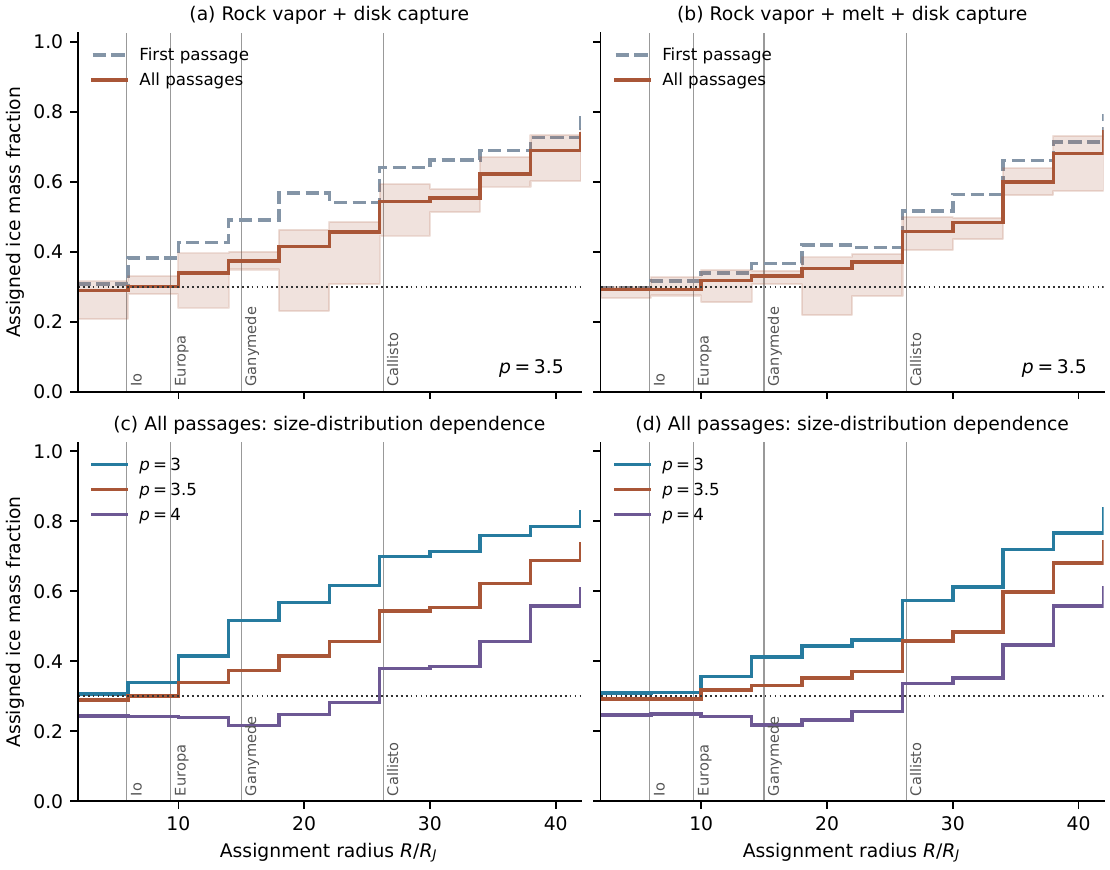}
\caption{Jupiter's radially assigned disk composition for the same
30/70 ice--rock source and capture prescription as the Saturn
calculation. Left and right columns give the vapor-only and
retained-melt rock treatments, including surviving disk parents.
Upper panels compare first passages with all passages at \(p=3.5\);
shading spans the four source-scramble estimates for all passages.
Lower panels use \(p=3,3.5,4\). Each step is a ratio of component
masses in a \(4R_J\) annulus. The figure displays \(2\)--\(42R_J\)
from the full \(2\)--\(100R_J\) calculation. The dotted line marks
the source ice fraction. Vertical guides mark the present Galilean
satellite orbits; these do not specify their feeding radii. The ice
fraction counts delivered water in both solid and vapor form; water
vapor becomes available as ice upon subsequent condensation. Planetary
absorption is excluded.}\label{fig:jupiter-composition}
\end{figure*}

Figure~\ref{fig:jupiter-composition} shows the assigned ice fraction
across the Galilean region. For \(p=3.5\), Jupiter's disk receives
0.25 of the incident ice mass and 0.11 or 0.13 of the incident rock
mass for vapor-only rock or retained melt, respectively. The combined
disk inventory has ice fractions of 0.49 and 0.45, exceeding the
0.30 source fraction. The corresponding values are 0.37 and 0.33 in
\(14\)--\(18R_J\), and 0.54 and 0.46 in \(26\)--\(30R_J\).
Across the four independent source scrambles, the \(14\)--\(18R_J\)
fractions span 0.35--0.40 and 0.31--0.35; the \(26\)--\(30R_J\)
fractions span 0.45--0.59 and 0.40--0.50. These are sampling ranges,
not confidence intervals. The whole-disk ranges are 0.48--0.50 and
0.45--0.47. Material supplied beyond Callisto's present orbit is still icier.

The Jovian background temperature is about 610 K at Io, 380 K at
Europa, 240 K at Ganymede, and 140 K at Callisto. The delivered water
inventory therefore includes vapor in the warm inner disk. Its
incorporation as solid ice involves subsequent cooling and condensation,
as in the ablation--recondensation picture of \citetalias{mosqueira2010}.
The assigned radial profiles record the delivery locations; later
transport of vapor and condensed material connects them to satellite
formation. A condensation front depends on the water partial pressure
as well as temperature, so it is not assigned a universal temperature
in this calculation.

Repeated passages add rock and lower the ice fraction relative to
the first-pass deposit, while retaining the outward increase across
the Galilean region. Reweighting the same histories at \(p=3\) and 4
changes the level and shape of the profile. For \(p=4\), the deposit
near Ganymede lies below the source ice fraction, while the outer
Galilean region remains enriched.
The \(\Theta=25\) comparison illustrates the competing effects of
encounter speed (Table~\ref{tab:jupiter-focusing}). The disk surface
density and temperature are identical in these runs, so their different
outcomes arise from the incoming population. Faster motion
increases aerodynamic heating at a given gas density, but the faster
incoming population also returns less often and retains less material.
At \(p=3.5\), water delivery falls from 0.25 to 0.20 of the incident
water mass, whereas rock delivery falls from 0.11 to 0.059 for vapor
loss alone, or from 0.13 to 0.076 with retained melt. Rock return
probabilities fall from 0.55/0.54 to 0.28/0.27. Ice supplies most of
its delivered mass on the first passage; rock benefits more from
repeated passages, whose suppression leaves more rock to escape.
Consequently, the bulk water fraction rises to 0.59/0.53 even though
both component delivery efficiencies decrease. Local fractions do
not all increase: with retained melt, the \(26\)--\(30R_J\) value
falls from 0.46 to 0.43. Both focusing choices nevertheless preserve
the increase from the Ganymede annulus to the Callisto annulus for
all three size-distribution indices.

The calculation establishes an ice-enriched Jovian disk inventory
from the rock-rich source. Migration and growth connect this
inventory to satellite compositions (Section~\ref{what-the-calculation-establishes}).

\subsection{Bulk ice enrichment}\label{bulk-ice-enrichment}

Table~\ref{tab:bulk-ice-enrichment} compares the mass-weighted ice
fractions of the full disk inventories for Saturn and Jupiter. Both
calculations use the canonical size-distribution index \(p=3.5\),
1\,m--1\,km fragments, a source containing 30\% ice by mass,
pressure-supported gas, \(\Theta=100\), and the same capture prescription. The totals
include delivery over all passages through \(2\)--\(100R_S\) for Saturn
and \(2\)--\(100R_J\) for Jupiter, extending to about \(0.1R_H\)
in each system. Each fraction is the total ice mass
delivered divided by the total ice-plus-rock mass delivered.

\begin{table}[t!]
\centering
\caption{Whole-disk ice mass fractions for a source ice fraction of
0.30, \(p=3.5\), and \(\Theta=100\). Both rock treatments include surviving disk-captured
remnants; the second also includes retained shed melt. Planetary
absorption is excluded.\label{tab:bulk-ice-enrichment}}
\begin{tabular}{@{}lcc@{}}
\toprule
Rock treatment & Saturn & Jupiter\\
\midrule
Vapor loss only & 0.45 & 0.49\\
Vapor loss and retained melt & 0.44 & 0.45\\
\bottomrule
\end{tabular}
\end{table}

Table~\ref{tab:bulk-delivery-channels} separates the delivery into vapor,
retained shed melt, and surviving captured remnants. Each entry is
normalized to the incident encounter mass of that material, before
applying the 30/70 source weights. Multiplying the ice total by 0.30
and the selected rock total by 0.70 gives the two masses used to
calculate each ice fraction in Table~\ref{tab:bulk-ice-enrichment}.

\begin{table*}[p]
\centering
\caption{Whole-disk delivery channels for the canonical \(p=3.5\)
population at \(\Theta=100\), per unit incident encounter mass of each material. All
passages are included. Captured remnants contribute their surviving
mass; planetary absorption is excluded. Totals and channels are rounded
independently to two significant figures.\label{tab:bulk-delivery-channels}}
\begin{tabular}{@{}llcccc@{}}
\toprule
Planet & Material treatment & Vapor & Retained melt & Captured remnant & Total\\
\midrule
Saturn & Ice & 0.25 & 0 & 0.021 & 0.27\\
 & Rock, vapor loss only & 0.10 & 0 & 0.043 & 0.14\\
 & Rock, vapor loss and retained melt & 0.013 & 0.11 & 0.031 & 0.15\\
\midrule
Jupiter & Ice & 0.24 & 0 & 0.0019 & 0.25\\
 & Rock, vapor loss only & 0.10 & 0 & 0.0093 & 0.11\\
 & Rock, vapor loss and retained melt & 0.011 & 0.11 & 0.0060 & 0.13\\
\bottomrule
\end{tabular}
\end{table*}

The integrated thermal contributions are similar: ice vapor delivery is
0.25 for Saturn and 0.24 for Jupiter, rock vapor delivery in the
vapor-loss-only treatment is 0.10 for both, and retained rock melt
contributes 0.11 for both. Saturn receives more surviving remnant mass.
Captured remnants supply 7.7\% of its delivered ice, compared with
0.78\% for Jupiter, but their share of delivered rock is larger still:
30\% or 20\% for Saturn, compared with 8.4\% or 4.7\% for Jupiter,
for vapor loss alone or vapor loss with retained melt, respectively.
Somewhat counterintuitively, Saturn's larger remnant contribution
increases rock delivery proportionally more than ice delivery and
accounts for its slightly lower bulk ice fractions. In both disks,
the total delivery efficiency for ice remains about twice that for
rock, producing similar bulk enrichment from the common source.

Differential ablation and capture therefore produce similar bulk ice
enrichment in both disks from the same rock-rich source. These
integrated inventories accompany the radial composition profiles in
Figures~\ref{fig:orbital-composition} and~\ref{fig:jupiter-composition};
the local annular ice fractions describe where that material is supplied.
The bulk enrichment and radial compositions reflect the combined
effects of differential ablation and remnant capture and are not solely
determined by the gas surface density of the circumplanetary disk.
In particular, the capture efficiency of rocky remnants also depends
on their incoming trajectories and velocities as they enter the disk,
which can offset the effect of a higher gas surface density.

\section{Discussion}\label{discussion}

\subsection{What the calculation establishes}\label{what-the-calculation-establishes}

The collisional study of \citet{mosqueira2026a} sought to explain the
observed satellite compositions through collisions within the Kronian
nebula. It showed that enriching the circumplanetary disk in ice was
easy to accomplish. The difficulty was producing a distant,
ice-rich Iapetus with a radius of about 700 km. Its substantial mass
and distant orbit make supplying enough ice-rich material challenging.
Here we examine how differential ablation and remnant capture can address 
this challenge.

Differential ablation and capture can supply that material while also
accounting for the bulk ice enrichment of the Jovian and Kronian
satellite systems from a rock-rich heliocentric source. From a source
of 30\% ice, the fixed radial distance reference
deposits material with ice fractions of 0.50--0.66 in the inner disk and
nearly pure ice farther out, and one incident normalization supplies
satellite-scale masses in both illustrative radial intervals. In the
orbital calculation, the ice fractions are 0.42 and 0.35 in
\(16\)--\(24R_S\), and 0.94 in \(56\)--\(64R_S\), for the vapor-only
and retained-melt rock treatments; repeated passages carry refractory
material farther inward while the outer deposit remains ice-rich. The
outward increase over the radial scale separating Titan and Iapetus
persists for both rock treatments and all three size-distribution
indices.

The Jupiter calculation applies the same source and capture prescription
to the Jovian disk. Its \(p=3.5\) disk inventory is enriched from
30\% ice to 49\% or 45\%, and the assigned ice fraction increases
outward across the Galilean region. At the common \(\Theta=100\),
differential ablation and remnant capture produce similar bulk ice
enrichment in both disks from the same rock-rich source
(Section~\ref{bulk-ice-enrichment}; Table~\ref{tab:bulk-delivery-channels}).
Saturn's larger remnant contribution
increases rock delivery proportionally more than ice delivery and
accounts for its slightly lower bulk ice fractions. In both disks,
the total delivery efficiency for ice remains about twice that for rock.

The bulk enrichment and radial compositions reflect the combined
effects of differential ablation and remnant capture. Gas surface
density alone does not determine the resulting composition: the capture
efficiency of rocky remnants also depends on their incoming trajectories
and velocities, which can offset the effect of a higher gas surface
density. Gas surface density affects ablation, 
while incoming trajectories and velocities affect remnant capture. 
Their combined effects yield similar ice-to-rock delivery efficiencies 
at the two planets.
The outer-disk Saturn calculation supplies both the ice-rich composition and
the mass needed for an Iapetus-sized satellite. The outward gradient is
a result for material delivered to the disk.

It is important to note that the delivery profiles 
describe the material added to the disk. They
are not predictions of individual satellite compositions. Satellitesimals
drift and growing satellites migrate inward through gas drag and disk
torques \citep{tanaka2002,mosqueira2026pds,mosqueira2026stall}, carrying
material delivered over a range of radii. Because the delivered ice fraction
increases outward, this inward transport raises an inner satellite's ice
fraction above that of its local annulus. The connection
between
inward migration and satellite composition will be the subject of a
forthcoming study.

\subsection{PDS~70~c: an observable analogue of the Iapetus region}\label{pds70-analogy}

\citet{mosqueira2026pds} connects the PDS~70 system to the formation of
Jupiter, Saturn, and their satellites. The two neighboring giant
planets share a circumstellar gap, while PDS~70~c retains a persistent
circumplanetary dust reservoir. That study finds that the inferred
emitting scale overlaps the circularization scale of late-stage,
gap-fed inflow, supporting an outer satellite-forming disk extending
to about \(0.1R_H\). The observed dust is consistent with a long-lived
reservoir in which
collisions and continued planetesimal-fragment delivery replenish
small grains. These results provide an additional setting for the
interaction between incoming solids and circumplanetary gas examined
here.

In particular, the outer disk of PDS~70~c occupies the analogous
satellite-forming region in an extended circumplanetary environment.
Ice enrichment could therefore be an observable property of the outer
disk of PDS~70~c. Continued delivery and processing of differentiated
fragments could supply icy grains, including ice recondensed from
ablated vapor, to its cold outer reservoir. A measurement of the
ice-to-refractory ratio there, compared with that of the incoming
circumstellar material, would directly address the compositional 
separation studied here.

\section{Conclusions}\label{conclusions}

\begin{itemize}
\item Differential ablation and capture of icy and rocky fragments
from differentiated outer Solar System bodies can explain the bulk ice
enrichment of the Jovian and Kronian satellite systems. They also supply
the substantial ice-rich mass needed in the outer Kronian disk to form
a distant satellite as large as Iapetus, with a radius of about 700 km.

\item For 1\,m--1\,km fragments with \(p=3.5\) from a source containing
30\% ice and 70\% rock, the fixed radial distance reference calculation
deposits ice fractions of 0.66 (delivery of vapor-only
rock) or 0.50 (retained melt) at \(20R_S\). These correspond to
ice-to-rock enrichment factors of 4.5 and 2.4, and the deposit is nearly
pure ice at \(60R_S\). Differential thermal ablation drives this
compositional selection: sublimation supplies 99\% of the ice deposited
at \(60R_S\), and shed
melt supplies 79\% of the rock deposited at \(20R_S\) in the
retained-melt treatment.

\item With one incident normalization, the same calculation supplies at
least one Titan mass between \(10\) and \(32R_S\) and 1.2 Iapetus masses
between \(60\) and \(63R_S\).

\item Surviving remnants return. In the Hill-entry ensemble with
pressure-supported gas, 70\% of vapor-only rocky objects that leave
the disk after one passage return. Disk capture or Saturn absorption follows for
97\% of returning vapor-only rock and 95\% with melt shedding.
Disk delivery rises from 0.058 to 0.17 of first-crosser mass with
vapor-only rock, and from 0.084 to 0.18 with retained melt.

\item The orbital \(p=3.5\) delivery has ice fractions of 0.42 or 0.35
in \(16\)--\(24R_S\), and 0.94 in \(56\)--\(64R_S\). This outward
gradient in the delivered material persists for both rock treatments, all three size-distribution indices, and
both tested vertical extents (Appendix~\ref{a.11.-vertical-extent}).
Local inner-disk enrichment above the source is not universal: the
\(p=4\) Titan-region deposit is below the source ice fraction.
For comparison, the Jovian deposit at \(26\)--\(30R_J\) has ice
fractions of 0.54 or 0.46 at \(\Theta=100\), well below the
Iapetus-region value. Saturn's low gas column near Iapetus permits
ice ablation while delivering little rocky material through either
ablation or capture. The outer-column test in
Section~\ref{capture-of-rocky-remnants-at-higher-outer-disk-columns}
shows that increasing the gas surface density captures more rocky
remnants and lowers the ice fraction. This helps explain the stronger
outer enrichment in the Kronian calculation.

\item Differential ablation and remnant capture produce similar bulk ice
enrichment in both disks from the same rock-rich source. Jupiter's
higher adopted outer-disk surface densities alone do not determine
whether its delivered material is more or less icy than Saturn's.
The capture efficiency of rocky remnants also depends on their incoming
trajectories and velocities, which can offset the effect of a higher
gas surface density.

The paired values below correspond to rock vapor loss alone or vapor
loss with retained shed melt, respectively. Both treatments include
captured remnants and delivered water vapor whose subsequent
condensation supplies solid ice. At the common focusing parameter
\(\Theta=100\) and canonical
size-distribution index \(p=3.5\), the disks extending to about
\(0.1R_H\) have whole-disk ice fractions of 0.45 or 0.44 for Saturn
and 0.49 or 0.45 for Jupiter, compared with 0.30 in the source
(Table~\ref{tab:bulk-ice-enrichment}). Counterintuitively,
Saturn's larger remnant contribution increases rock delivery
proportionally more than ice delivery, accounting for its slightly
lower bulk ice fractions. In both disks, ice is delivered about twice
as efficiently as rock.

A separate comparison holds the Jovian disk fixed and changes
\(\Theta\) from 100 to 25. The faster incoming population retains
water supplied by ice ablation while allowing more rock to escape.
Reduced returns suppress rock delivery more strongly than water
delivery, raising the bulk ice fractions to 0.59 or 0.53 even though
less water is delivered. This illustrates how incoming velocities
can change the balance between ice and rock delivery even at the same
gas surface density. At the level of accuracy of the
present calculation, bulk ice enrichment is expected for both planets.
Establishing more precise differences between their bulk compositions
requires more detailed modeling of the disks and incoming populations.
Disk capture, satellitesimal migration, and satellite 
incorporation are separate issues requiring further modeling.

\end{itemize}

\begin{acknowledgments}
The author acknowledges assistance from ChatGPT (OpenAI) with model development,
numerical implementation, consistency checks,
and manuscript preparation. Claude Opus 5.5 (Anthropic) provided editorial assistance.\end{acknowledgments}

\needspace{18\baselineskip}
\appendix
\setcounter{equation}{0}\renewcommand{\theequation}{A\arabic{equation}}
\renewcommand{\theHequation}{A.\arabic{equation}}
\setcounter{figure}{0}\setcounter{table}{0}
\renewcommand{\thefigure}{A\arabic{figure}}\renewcommand{\thetable}{A\arabic{table}}
\renewcommand{\theHfigure}{A.\arabic{figure}}\renewcommand{\theHtable}{A.\arabic{table}}
\section{Model specification and numerical
checks}\label{appendix-a.-model-specification-and-numerical-checks}

\subsection{Continuous disk
interpolation}\label{a.1.-continuous-disk-interpolation}

Empirical-law coefficients, adopted input measurements,
mass-normalization constants, and numerical-check limits retain
additional digits in this appendix for reproducibility. Reported
physical outcomes are otherwise rounded to two significant figures.

Let \(\xi=(X-20)/38\) be a dimensionless interpolation coordinate
clipped to \([0,1]\), and \(W=6\xi^5-15\xi^4+10\xi^3\) its blending
weight. Then

\begin{equation}
\Sigma_g=\Sigma_{\rm in}
\left(\frac{\Sigma_{\rm out}}{\Sigma_{\rm in}}\right)^W.
\label{eq:legacy-A1}
\end{equation}

This explicit interpolation joins the published branch values
continuously and matches their endpoint slopes. The fixed radial distance
reference follows vertical motion.
Gas rotation in the orbital extension is prescribed separately, as
specified in Section~\ref{orbital-continuation-and-repeated-disk-passages}.

\subsection{Material laws and effective removal
energies}\label{a.2.-material-laws-and-effective-removal-energies}

In the following material laws, \(T\) denotes surface temperature in
kelvin and \(P_i\) the saturation vapor pressure over ice. Let
\(a_n,b_n\) be fitted dimensionless coefficients, with \(n\) indexing
their terms. Define the reduced temperature
\(\theta=T/(273.16\,\mathrm K)\). Below the triple point, the pressure
follows the \citet{iapws2011} relation, valid from 50 to 273.16 K:

\begin{equation}
P_i=611.657\exp\!\left[\theta^{-1}
\sum_{n=1}^3a_n\theta^{b_n}\right]\ \mathrm{Pa},
\label{eq:legacy-A2}
\end{equation}

with \(a_n=(-21.2144006,\,27.3203819,\,-6.10598130)\) and
\(b_n=(0.00333333333,\,1.20666667,\,1.70333333)\). Above 273.16 K, let
\(P_l\) be the saturation vapor pressure over liquid water and
\(\eta=1-T/(647.096\,\mathrm K)\). With fitted dimensionless
coefficients \(c_n,d_n\), the code uses the branch given by
\citet{wagner1993}, also tabulated by \citet{dangelo2015}:

\begin{equation}
\ln\frac{P_l}{22.064\times10^6\ \mathrm{Pa}}
=\frac{647.096}{T}\sum_{n=1}^6c_n\eta^{d_n}.
\label{eq:legacy-A3}
\end{equation}

The coefficient values are:

\[
\begin{aligned}
c_n={}&(-7.85951783,\,1.84408259,\,-11.7866497,\\
      &22.6807411,\,-15.9618719,\,1.80122502),\\
d_n={}&(1,1.5,3,3.5,4,7.5).
\end{aligned}
\]

This avoids extrapolating the ice law above its range. The largest
sampled ice-body surface temperature in the reference ensemble is
approximately 340 K.

For refractory saturation vapor pressure \(P_r\), the rock comparison
uses the effective quartz pressure fit from \citet{dangelo2015},

\begin{equation}
P_r=0.1\exp\!\left(31.82319964-
\frac{46071.4304}{T+58.883}\right)\ \mathrm{Pa}.
\label{eq:legacy-A4}
\end{equation}

The effective molar masses are 0.01801528 and
\(0.0601\ \mathrm{kg\,mol^{-1}}\) for water and the refractory vapor
proxy. We use this single refractory pressure fit throughout each
trajectory. Its
extremely small low-temperature fluxes are interpreted as negligible.

Let \(c_i(T)\) be the specific heat capacity of ice and \(h_i(T)\) its
sensible enthalpy increase from 100 K; \(T'\) below is the integration
temperature. For the steady receding-surface energy closure,
\(Q_{v,i}=2.83\times10^6+h_i(T)\) J kg\(^{-1}\). Below 273.16 K we take
\(c_i(T)=7.49T+90\ \mathrm{J\,kg^{-1}\,K^{-1}}\) and
\(h_i(T)=\int_{100\,\mathrm K}^{T}c_i(T')\,dT'\). Above it, the sensible
term continues as \(h_i(273.16)+4180(T-273.16)\) J kg\(^{-1}\). The
constant phase energy is used as an effective solid-to-vapor cost; where
a liquid layer intervenes it already includes fusion plus evaporation.
The energy closure thus combines a constant phase-change energy with
the stated temperature-dependent sensible enthalpy, counting fusion once.

For rock with specific heat capacity \(c_r\),
\(Q_{v,r}=8.08\times10^6+c_r(T-100)\) J kg\(^{-1}\), with
\(c_r=947.06\ \mathrm{J\,kg^{-1}\,K^{-1}}\). This effective heat
capacity is chosen so that sensible heating to 1800 K plus the
\(1.43\times10^6\ \mathrm{J\,kg^{-1}}\) fusion energy reproduces
\citetalias{mosqueira2010}'s \(3.04\times10^6\ \mathrm{J\,kg^{-1}}\)
melting budget. The baseline phase energies, heat capacities, density,
and pressure fit define the reference material model. The same values
are used in every comparison; none is adjusted to match an observed
composition.

\subsection{Numerical integration and
conservation}\label{a.3.-numerical-integration-and-conservation}

The code integrates equation~\eqref{eq:legacy-4} in dimensionless height \(\zeta=z/H\),
normal velocity \(V=u/c_s\), and time \(\tau=\Omega_Kt\), using
\(\ln(s/s_0)\) for the evolving size and cumulative vapor and melt
fractions as additional variables. A DOP853 integrator locates exit,
confinement, and a 1 cm radius cutoff. The residual mass at that cutoff
is credited to ablation; it is at most \(10^{-6}\) of the mass of the
smallest initial fragment.

Surface solutions are obtained by monotone interpolation of the combined
cooling function in equation~\eqref{eq:legacy-8}, with 14,001 base temperature nodes
plus phase endpoints. Checked fractional surface-energy residuals are
below \(4.5\times10^{-6}\). Across the reference size-integrated
ensemble, individual-trajectory component-mass residuals are below
\(2.7\times10^{-6}\) of initial mass.

Size integration uses 24-point Gauss--Legendre quadrature in \(\ln s_0\)
on each side of the complete-delivery boundary, which is located by
bisection. A test doubling the quadrature order and tightening the
trajectory tolerance at \(20\), \(40\), \(60\), and \(100R_S\) gives a
maximum component-efficiency change of \(4.36\times10^{-6}\). A repeat
with Python 3.12.14, NumPy 2.5.3, and SciPy 1.18.1 gives
\(1.50\times10^{-6}\) between fresh base and refined integrations, and
\(4.80\times10^{-6}\) between the stored production values and fresh
refined results. Thus \(5\times10^{-6}\) in absolute fraction covers the
tested stored-to-recomputed differences. The radial calculation contains
70 annuli. These checks measure integration error for the adopted
heat-transfer and material laws.

The figures specify the reference population or the single input varied
in each comparison.

\subsection{Composition references and the Saturn-disk
overlays}\label{a.4.-composition-references-and-the-saturn-disk-overlays}

All composition comparisons assume negligible porosity and distinguish
refractory material from the water-rich component. The refractory
fraction includes silicates and metal; an entire hydrated core must not
be counted as dry rock. The water category includes solid and liquid
phases and, where included by the cited model, water bound in hydrated
material. Sohl et al.'s Titan water-rich layers also contain minor
ammonia. The figure compares these specified interior models.

Let \(V_B\) be the volume of body
\(B\), with \(B=T,I\) denoting Titan and Iapetus, and \(dm\) a mass
element. Interior pressure and temperature are \(P\) and \(T\);
water-rich density \(\rho_w\) also depends on phase, and refractory
density \(\rho_r\) on composition. Volume accounting gives:

\begin{equation}
\begin{aligned}
V_B&=\int_{\rm water}\frac{dm}{\rho_w(P,T,{\rm phase})}\\
&\quad+\int_{\rm refractory}\frac{dm}{\rho_r(P,T,{\rm composition})},\\
B&\in\{T,I\}.
\end{aligned}
\label{eq:legacy-18}
\end{equation}

Table~\ref{tab:A1} lists every Figure~\ref{fig:observational-context} point. A published interval is retained as
a model-family or reference-model interval with its original meaning. Midpoints of
published ranges are used only to place symbols. For the four new compact
reference calculations, the interval combines the measurement or shape
bounds described in Appendix~\ref{a.7.-size-data-and-the-compact-reference-calculations} with a 100--250 K water-shell
temperature range at fixed refractory mineralogy. Triton's 0.65--0.70
interval is a stated literature estimate. Each range corresponds to
the mineralogy, thermal profile, and interior structure of its model.

\begin{table*}[t!]
\centering
\caption{Adopted zero-porosity composition references for Figure~\ref{fig:observational-context}. Water-rich
mass fractions are the complements of the listed refractory fractions.
Calculated fractions are rounded to two significant figures; published
values keep their quoted precision. Published model ranges, the Triton
literature estimate, and compact-reference sensitivities have different
meanings (Appendix~\ref{a.4.-composition-references-and-the-saturn-disk-overlays}).\label{tab:A1}}
\small\setlength{\tabcolsep}{4pt}\renewcommand{\arraystretch}{1.15}
\begin{tabular}{@{}>{\raggedright\arraybackslash}p{(\linewidth - 8\tabcolsep)*\real{0.12000000}}>{\raggedright\arraybackslash}p{(\linewidth - 8\tabcolsep)*\real{0.11000000}}>{\raggedright\arraybackslash}p{(\linewidth - 8\tabcolsep)*\real{0.09000000}}>{\raggedright\arraybackslash}p{(\linewidth - 8\tabcolsep)*\real{0.17000000}}>{\raggedright\arraybackslash}p{(\linewidth - 8\tabcolsep)*\real{0.51000000}}@{}}
\toprule
Body & Radius (km) & Rock + metal fraction & Displayed interval & Reference basis\\
\midrule
Quaoar & 547 & 0.65 & 0.60--0.70 & Compact reference\\
Charon & 606 & 0.59 & 0.575--0.605 & \citet{mckinnon2017}\\
Iapetus & 734 & 0.20 & 0.18--0.23 & Compact reference\\
Haumea & 798 & 0.70 & 0.65--0.73 & Compact reference; shape dependence\\
Eris & 1163 & 0.87 & 0.85--0.88 & Compact reference\\
Pluto & 1188 & 0.655 & 0.650--0.660 & \citet{mckinnon2017}\\
Triton & 1353 & 0.68 & 0.65--0.70 & \citet{cioria2022}\\
Callisto & 2410 & 0.48 & 0.45--0.51 & \citet{kuskov2005}\\
Titan & 2575 & 0.55 & 0.50--0.60 & \citet{sohl2003}\\
Ganymede & 2631 & 0.53 & 0.52--0.54 & \citet{kuskov2001}\\
\bottomrule
\end{tabular}
\end{table*}

Titan, Ganymede, and Callisto use interior calculations with
pressure-dependent water phases, rather than the low-pressure relation
used previously for the satellite overlays.
\citet[paragraph 27 and Figure 5]{sohl2003} report 50--60 wt\% anhydrous
rock plus iron. Their layered Titan models include high-pressure ice but
adopt constant densities within each layer. \citet{kuskov2001} obtain
46--48 wt\% water for Ganymede. \citet{kuskov2005} obtain 49--55 wt\%
water for partially differentiated Callisto; the refractory complement
counts rock in both the core and mixed interior. The interval widths
reflect the selected interior-model families.

\citet[Figure 2]{mckinnon2017} report dry refractory fractions
\(0.655\pm0.005\) and \(0.590\pm0.015\) for their Pluto and Charon
reference interiors. Their quoted fractions assign hydration water in
the adopted PF2 material to the water budget. For Triton,
\citet[Introduction]{cioria2022} explicitly state an estimated
rock-plus-metal mass fraction of 65--70\%, citing McKinnon and Kirk
(2014). We use that stated interval, a literature synthesis, rather than
an estimate constructed from the Pluto comparison. The complementary
water-rich fraction is 30--35\%.

Figure~\ref{fig:gradient} uses only Titan and Iapetus. Denote their reference water-rich
mass fractions by \(f_{i,T}\) and \(f_{i,I}\). Titan has
\(f_{i,T}=0.45\) with the selected range 0.40--0.50; Iapetus has
\(f_{i,I}=0.80\) with the compact reference range 0.77--0.82. Figure~\ref{fig:column-comparison}
uses that same Iapetus interval. There are no density-only error bars
derived from a fixed \(0.92\ \mathrm{g\,cm^{-3}}\) water density for
Titan, and no adjustable porosity in these comparisons. Ganymede and
Callisto appear only in Figure~\ref{fig:observational-context}, not as additional locations in
Saturn's disk.

\subsection{Independent checks and mass
integration}\label{a.5.-independent-checks-and-mass-integration}

The source code uses SI units. Let \(A_p\) be projected area and \(A_s\)
full surface area. The production and audit solvers both use
\(A_p=\pi s^2\) in drag and aerodynamic power and \(A_s=4\pi s^2\) in
surface losses. Consequently, \(q_{\rm aero}=C_H\rho_g|u|^3/8\), and
mass loss changes the same radius appearing in the drag acceleration.
The vapor and melt terms consume one common energy budget. The confined-remnant
contribution is the surviving mass at confinement, not the original
fragment mass.

An independent solver integrates 18 representative crossings, with
direct root solves of the surface balance in place of the production
lookup table. It also integrates the dissipated vertical specific
energy. All confinement/exit classifications agree. The largest difference
in a component mass fraction between the direct and production solutions
is \(1.42\times10^{-7}\); the independent mass budget residual is below
\(1.2\times10^{-10}\) and the relative vertical-energy residual below
\(7.6\times10^{-11}\). Linear-size and logarithmic-size evaluations of
the source-mass normalization agree to \(4.5\times10^{-16}\) in the
tested cases. These are checks of the fixed radial distance crossing calculation.

The mass calculation evaluates \(p=3\), 3.5, and 4 on 139 radial nodes.
Each component is integrated with 24-point Gauss--Legendre quadrature in
\(\ln s_0\) on both sides of the complete-delivery boundary. The
per-radius mass kernels are interpolated with shape-preserving piecewise
cubics and integrated analytically over each interpolating segment. The
local inventory calculation uses those efficiencies; it changes no
crossing or ablation parameter.

For reproducibility, the inventory calculation uses
\(M_T=1.34518\times10^{23}\) kg and \(M_I=1.80566\times10^{21}\) kg at
the displayed precision, retaining the unrounded JPL-based values in the
calculation. Let \(M_{\rm pass}\) be the distinct incident fragment mass
over \(8\leq X\leq100\), \(M_\oplus\) Earth's mass, and \(I\) the
dimensionless area-weighted focusing integral. The numerical source
normalization is equivalent to

\begin{equation}
\Sigma_\star=\frac{M_{\rm pass}}{2\pi R_S^2 I},\qquad
I=\int_8^{100}Xg(X)\,dX=23368,
\label{eq:legacy-A5}
\end{equation}

with \(M_{\rm pass}=0.5M_\oplus\), giving
\(\Sigma_\star=564.93455\ \mathrm{g\,cm^{-2}}\). This equality records
the normalization of the existing dataset. The finite-interval
calculation in equation~\eqref{eq:legacy-16} uses \(\Sigma_\star\) directly; changing
the remote computational boundary at fixed \(\Sigma_\star\) does not
change the local deposited mass. \(M_{\rm pass}\) counts distinct
incoming 1--1000 m fragments in that bookkeeping domain, not all
planetesimals in the solar nebula.

Figures~\ref{fig:size-titan}--\ref{fig:powerlaw} show the crossing and power-law results. Figures~\ref{fig:local-titan}--\ref{fig:local-iapetus}
integrate the same \(p=3.5\) mass kernels over the local intervals in
Table~\ref{tab:3}. No satellite composition or global outer-disk inventory is used
to fit the incident amplitude.

\subsection{Comparison of the old and revised melting
laws}\label{a.6.-comparison-of-the-old-and-revised-melting-laws}

Let \(J_{r,\rm old}\) be the positive rock-removal flux per spherical
surface area in \citetalias{mosqueira2010}'s equation (5),
\(q_{\rm abs,old}\) its absorbed background flux, \(T_0\) its background
temperature, and \(E_{0m}\) its effective melting energy per unit mass.
Writing \([y]_+=\max(y,0)\) for the positive part of a real argument
\(y\), the old law is

\begin{equation}
J_{r,\rm old}=\frac{
\left[q_{\rm aero}+q_{\rm abs,old}-\epsilon\sigma_{\rm SB}T_m^4\right]_+}{E_{0m}},
\qquad [y]_+=\max(y,0).
\label{eq:legacy-A6}
\end{equation}

The corresponding no-ablation temperature diagnostic determined whether
this branch was active. All net power above the prescribed radiative
loss was assigned to one effective ablation flux using the melting
energy \(E_{0m}\). In \citetalias{mosqueira2010}'s printed equation,
\(q_{\rm abs,old}=\sigma_{\rm SB}T_0^4\); the present gray-background
convention uses absorbed flux
\(q_{\rm abs}=\epsilon\sigma_{\rm SB}T_g^4\). This
background-normalization difference is separate from the revised
mass-loss prescription. \citetalias{mosqueira2010} already recognized
that melt viscosity could affect removal, but did not solve liquid flow
or breakup (\citetalias{mosqueira2010}, Section 3.1).

To isolate the prescription change, compare the two laws at the same
absorbed background flux \(q_{\rm abs}\) and removal energy. Denote the
available surface power per unit area at melting by \(\mathcal H_m\):

\begin{equation}
\mathcal H_m=q_{\rm aero}+q_{\rm abs}-\epsilon\sigma_{\rm SB}T_m^4.
\label{eq:legacy-A7}
\end{equation}

The old effective law assigns \(\mathcal H_m/Q_{m,r}\) to mass loss when
\(\mathcal H_m>0\). Denote the revised total refractory removal flux by
\(J_{r,\rm new}=J_{v,r}+J_{m,r}\). The new melt branch requires the
stronger condition \(\mathcal H_m>Q_{v,r}(T_m)J_{v,r}(T_m)\), because
refractory vapor is already carrying energy away. On that active branch,

\begin{equation}
\begin{aligned}
J_{r,\rm new}&=J_{v,r}(T_m)+J_{m,r}\\
&=\frac{\mathcal H_m}{Q_{m,r}}
-\left[\frac{Q_{v,r}(T_m)}{Q_{m,r}}-1\right]J_{v,r}(T_m).
\end{aligned}
\label{eq:legacy-A8}
\end{equation}

Thus the change is not a second melting yield added to an unchanged
vapor yield. Vaporization spends more energy per unit mass and reduces
the power left for melt shedding. In the limit of negligible vapor loss,
the new melt branch recovers the old effective energy-limited form,
after matching the background convention and removal energy. At lower
heating, the new model permits vapor loss even when its melting branch
is inactive. The melting threshold itself is retained; what changes is
the treatment of vapor loss and its competition with melt removal.

For the adopted refractory proxy,
\(J_{v,r}(1800\,\mathrm K)\simeq0.091\ \mathrm{kg\,m^{-2}\,s^{-1}}\) and
\(Q_{v,r}(1800\,\mathrm K)\simeq9.7\times10^6\ \mathrm{J\,kg^{-1}}\).
With a 100 K background, vapor therefore uses
\(8.8\times10^5\ \mathrm{W\,m^{-2}}\) at melting. The aerodynamic flux
needed to activate melt shedding is approximately
\(1.2\times10^6\ \mathrm{W\,m^{-2}}\), compared with
\(3.0\times10^5\ \mathrm{W\,m^{-2}}\) for a radiation-only melting
threshold. Both thresholds follow from the stated material law.

An independent check evaluates the refractory law at \(T_m\), verifies
equation~\eqref{eq:legacy-A8}, and obtains the Section~\ref{size-integrated-delivery-channels} channel percentages from the
calculated channel totals. The no-vapor limit recovers the old law after
matching absorbed background flux and removal energy. This comparison
uses \citetalias{mosqueira2010}'s printed equation (5) and material
table.

The current calculation retains \(T_m\), \(Q_{m,r}\), the stated
fragment population, and the unit-retention assumption in equation~\eqref{eq:legacy-12a}. Changing the vapor--melt allocation is distinct from changing the
assumed fate of the shed material.

\subsection{Size data and the compact reference
calculations}\label{a.7.-size-data-and-the-compact-reference-calculations}

The linear size coordinate is volume-equivalent mean radius, not mass or
a projected major axis. The ten bodies in Table~\ref{tab:A1} are all at least as
large as Quaoar. Ganymede, Callisto, Titan, Iapetus, and Triton use the
JPL SSD radii. Pluto and Charon use the New Horizons mean radii. Quaoar uses \citeauthor{margoti2026}'s
\citeyearpar{margoti2026} oblate solution, with volume-equivalent radius
\(547.2\pm2.3\) km and density \(1.760\pm0.109\ \mathrm{g\,cm^{-3}}\).
Haumea's nominal \(798\) km volume-equivalent radius follows from its
measured mass and the \(1.885\ \mathrm{g\,cm^{-3}}\) occultation
solution \citep{ortiz2017}; no longest-axis radius is substituted.

Eris must be distinguished from the volume-averaged Eris--Dysnomia
system. We combine the individual radius \(1163\pm6\) km
\citep{sicardy2011} with system mass \((1.6466\pm0.0085)\times10^{22}\)
kg \citep{holler2021}. The nominal primary-mass approximation gives
Eris's mean density, \(\rho_E\simeq2.5\ \mathrm{g\,cm^{-3}}\). The
conditional density range also includes satellite-to-primary mass ratios
from zero to 0.0085, the published \(1\sigma\) upper-limit scale in
\citet{brown2023}. This is not the earlier combined-volume density of
\(2.43\ \mathrm{g\,cm^{-3}}\); the displayed range combines the
specified radius and mass-ratio inputs.

For Iapetus, Quaoar, Haumea, and Eris we calculate an explicitly
specified compact reference: a dry olivine interior and an ice Ih shell,
each with pressure-dependent density. The nominal temperatures are 300 K
in the refractory component and 150 K in the shell. Let \(\rho(P,T)\) be
material density at pressure \(P\) and temperature \(T\), \(\rho_0(T)\)
its zero-pressure density, \(K_T(T)\) its zero-pressure isothermal bulk
modulus, and \(K'\) the dimensionless pressure derivative of that
modulus. Using the material constants tabulated by \citet{fortes2012},
the compression law is

\begin{equation}
\rho(P,T)=\rho_0(T)\left[1+\frac{K'P}{K_T(T)}\right]^{1/K'}.
\label{eq:legacy-A9}
\end{equation}

For ice Ih,
\(\rho_0(T)=934.31\exp[-3.348\times10^{-8}T^{2.559}/2.559]\ \mathrm{kg\,m^{-3}}\),
\(K_T=(10.995-0.004068T-2.051\times10^{-5}T^2)\) GPa, and \(K'=7\). For
the dry San Carlos olivine reference,
\(\rho_0(300\,\mathrm K)=3343\ \mathrm{kg\,m^{-3}}\),
\(K_T=(131.1-0.0223T)\) GPa, \(K'=3.8\), and thermal expansivity is
\(2.66\times10^{-5}\ \mathrm{K^{-1}}\). These are reference mineral
properties, not fitted fragment parameters.

In this interior calculation, \(r\) is distance from the body's center,
\(m(r)\) the enclosed mass, and \(\omega\) its spin angular frequency.
For each measured mass and radius, a shooting calculation adjusts
central pressure and core radius to satisfy \(dm/dr=4\pi r^2\rho\) and
\(dP/dr=-Gm\rho/r^2\), with zero surface pressure. Haumea uses a
volume-equivalent spherical average with the centrifugal term
\(+(2/3)\rho\omega^2r\) for its 3.9155 h rotation. The nominal rock fraction is the
integrated refractory mass divided by total mass. The temperature and
measurement/shape comparisons in Table~\ref{tab:A1} are conditional on this
mineralogy and layer arrangement. Haumea's interval includes Ortiz et
al.'s 5\% ring-light density alternative,
\(1.757\ \mathrm{g\,cm^{-3}}\), as well as the nominal
\(\pm0.080\ \mathrm{g\,cm^{-3}}\) density uncertainty. The interval
uses these density and temperature variations at fixed interior composition.

The largest ice-shell base pressure in these compact reference
calculations is 0.14 GPa. Their ice Ih treatment is therefore not
extended into the multi-GPa water interiors of Titan, Ganymede, or
Callisto; those use the cited published layered models including
high-pressure ice. Disabling compression recovers the analytical
two-density mass fraction to better than \(10^{-7}\). The calculation
solves the interior structure at the specified temperatures.

The composition comparison combines the adopted literature fractions and
four compact reference inversions. Figures~\ref{fig:observational-context}, \ref{fig:gradient}, and \ref{fig:column-comparison} use these values,
with the assumptions and sources specified above and in Table~\ref{tab:A1},
together with the deposition calculation. The Iapetus porosity scale
calculation in Section~\ref{observational-context} remains distinct from these zero-porosity
inversions. Neither this composition comparison nor adding Ganymede and
Callisto changes any incoming fragment property, ablation/capture
efficiency, or deposited-mass result.

\subsection{Orbital integration and conditional-probability
checks}\label{a.8.-orbital-integration-and-conditional-probability-checks}

\textbf{Integration variables and units.} The orbital calculation uses
the following length, time, and speed units:

\[
\text{length: }R_S,\qquad
\text{time: }\left(\frac{R_S^3}{GM_S}\right)^{1/2},\qquad
\text{speed: }\left(\frac{GM_S}{R_S}\right)^{1/2}.
\]

A DOP853 integrator evolves position, rotating-frame velocity,
logarithmic fragment radius, vapor and melt fractions, and the integrated
drag contribution to the Jacobi constant. Between gas encounters the
fragment radius \(s\) is constant. Within them, vapor and melt removal use the same lookup
tables as the reference solver. Isotropic recession is assumed: removed
material leaves with the parent's bulk velocity, so no recoil term is
introduced.

The solar gravitational parameter and conversion constants are

\[
\begin{aligned}
GM_\odot &= 1.32712440018\times10^{20}\ \mathrm{m^3\,s^{-2}},\\
1\ \mathrm{au} &= 149597870700\ \mathrm m,\\
1\ \mathrm{yr} &= 365.25\ \mathrm{days}.
\end{aligned}
\]

Saturn's parameters and the material constants are unchanged.

\textbf{Step control and boundary detection.} Incoming ballistic
integrations use a relative tolerance of
\(2\times10^{-10}\). The gas-interacting histories in all incoming ensembles
use \(2\times10^{-11}\). Let \(\Delta t_{\max}\) denote the maximum
allowed step. Its local value is

\[
\Delta t_{\max}=0.05\left(r^3/GM_S\right)^{1/2}.
\]

Gas boundaries are checked at eight subdivisions of each accepted
ballistic step and sixteen for gas-interacting histories, including
the controlled examples.
If an explicit Runge--Kutta trial stage gives a nonpositive squared
gas speed outside the physical gas domain, the step is retried from the
last accepted state at half its size. This prevents an invalid trial
state from terminating the integration; the physical gas law is unchanged.
The first crossing is root-located, and the force evaluation restarts
with the state continuous across the boundary. Checking only the step
endpoints can miss a short grazing exit and return; the interior checks
are therefore applied to both first-entry and subsequent-passage
integrations. The capture criterion is also checked within each step to
locate its first satisfaction.

\textbf{Conserved quantity without gas.} Particle energy and angular
momentum need not remain constant under stellar tides. The conserved
quantity for gas-free Hill motion is the Jacobi constant, denoted by
\(C_J\):

\begin{equation}
C_J=3n_p^2x^2-n_p^2z^2+\frac{2GM_S}{r}
-|\boldsymbol v_{\rm rot}|^2.
\label{eq:legacy-A10}
\end{equation}

Independent integrations without gas conserve this quantity. A separate
test also disables the solar terms and verifies conservation of the
two-body energy.

\textbf{Work balance with gas drag.} Drag changes the Jacobi constant.
Its expected rate of change is

\begin{equation}
\dot C_J=-2\boldsymbol v_{\rm rot}\cdot\boldsymbol a_D.
\label{eq:legacy-A11}
\end{equation}

\Needspace{12\baselineskip}

We integrate the right-hand side alongside the trajectory and compare
the accumulated work with the actual change in the Jacobi constant. Let
\(\delta_J(t)\) denote this dimensionless balance residual. Time is
measured from the start of the trajectory, and \(t'\) is the integration
variable:

\begin{equation}
\delta_J(t)=\frac{
C_J(t)-C_J(0)
+2\displaystyle\int_0^t\boldsymbol v_{\rm rot}(t')\cdot\boldsymbol a_D(t')\,dt'
}{GM_S/R_S}.
\label{eq:legacy-A12}
\end{equation}

The largest absolute residual over the primary incoming-ensemble histories is
below \(9.7\times10^{-11}\). The prescribed gas is an external momentum
and heat reservoir. Equation~\eqref{eq:legacy-8} separately closes the steady surface
energy budget; neither check assumes conservation of the particle's
mechanical energy alone. The corresponding mass-accounting checks are
given in Appendix~\ref{a.9.-radial-composition-mass-accounting-and-numerical-checks}.

\textbf{Independent integrations and numerical refinement.} An independent continuous-force integrator, without the production event
restarts, reproduces five representative outcomes with component
differences below \(3.5\times10^{-9}\). Eight independent incoming integrations
reproduce the first-entry or miss classifications.

An additional paired refinement of all 512 incoming geometries halves
the first-entry step and doubles the boundary sampling. Every entry/miss
classification is preserved. Orbit and deposition refinement for the
adopted capture criterion is reported in Appendix~\ref{a.9.-radial-composition-mass-accounting-and-numerical-checks}.

\textbf{Map from Sobol coordinates to the incoming state.} Let
\(U_1,U_2,U_3,U_4\) denote the four unit-interval Sobol coordinates. Let
\(\vartheta\) and \(\varphi\) be the polar and azimuthal angles of the
Hill-surface position, and \(\psi\) the angle of its tangential velocity
in the local tangent plane. The sampled variables are

\begin{equation}
\begin{aligned}
b_\infty^2&=U_1\left(R_{\max}^2+\frac{2GM_SR_{\max}}{v_\infty^2}\right),\\
\cos\vartheta&=2U_2-1,\qquad
\varphi=2\pi U_3,\qquad \psi=2\pi U_4.
\end{aligned}
\label{eq:legacy-A13}
\end{equation}

Define the spherical orthonormal basis at that position by

\[
\begin{aligned}
\boldsymbol e_r&=(\sin\vartheta\cos\varphi,\sin\vartheta\sin\varphi,\cos\vartheta),\\
\boldsymbol e_\vartheta&=(\cos\vartheta\cos\varphi,\cos\vartheta\sin\varphi,-\sin\vartheta),\\
\boldsymbol e_\varphi&=(-\sin\varphi,\cos\varphi,0).
\end{aligned}
\]

Let \(\boldsymbol v_{\rm in}\) denote the planet-centered inertial
velocity. Using the signed radial speed \(v_r\) and tangential speed
\(v_t\) specified in Section~\ref{incoming-ensemble-and-passage-probabilities}, the initial position and velocity are

\begin{equation}
\begin{aligned}
\boldsymbol r&=R_H\boldsymbol e_r,\\
\boldsymbol v_{\rm in}&=v_r\boldsymbol e_r+v_t(\cos\psi\boldsymbol e_\vartheta+\sin\psi\boldsymbol e_\varphi),\\
\boldsymbol v_{\rm rot}&=\boldsymbol v_{\rm in}-\boldsymbol n_p\times\boldsymbol r.
\end{aligned}
\label{eq:legacy-A14}
\end{equation}

Uniform polar cosine, azimuth, and tangent-plane angle give rotationally
invariant orientations; uniform polar angle would not.

\textbf{Encounter sampling and size resolution.} The incoming ensemble
uses four independently scrambled Sobol sequences with seeds 731--734
and 128 encounter geometries each. Successful entries are integrated
over 64 equal logarithmic size bins from 1 to 1000 m. Trajectories use
bin-center sizes, while number and mass weights are integrated exactly
over each bin.

The pilot uses the same first 64 geometries in each scramble and 32 size
bins. Holding those 256 encounter geometries fixed and doubling the size
resolution changes the cumulative component delivery efficiencies by
less than \(0.00068\) for size-distribution index 3.5. Variation among the
four independent source scrambles is reported in Section~\ref{first-encounters-returns-and-cumulative-capture} and Figure~\ref{fig:orbital-composition}. It measures finite sampling of the adopted encounter distribution.

\textbf{Physical endpoints and continuation.} The integration initially
uses a 100 yr numerical checkpoint. If a history exhausts its
computational step allowance, that allowance is doubled and the same
history is rerun without changing its physical or accuracy parameters. Any surviving history is resumed
from its full saved state, doubling the checkpoint until a physical
endpoint is reached. Position, velocity, size, accumulated mass loss,
and integrated drag work are preserved. All 81,024 gas-interacting
histories reach endpoints: 30,405 reach disk capture, 1,271 cross
the inner absorbing boundary, 680 fully ablate, and 48,668 exit at the
Hill boundary. Sixteen require continuation past the first
checkpoint, with the last ending at about \(1.7\times10^{2}\) yr. A
restart test reproduces uninterrupted endpoint classifications and mass
fractions to within \(2\times10^{-15}\). Simultaneous inner-boundary
absorption and gas-exit events give priority to absorption, with a
\(10^{-10}R_S\) positional tolerance. The inner absorbing sphere is at
\(2R_S\); outward crossing of \(R_H\) ends the modeled encounter.

The Keplerian-gas numerical control is reported separately in
Appendix~\ref{a.10.-gas-rotation-and-physical-boundary-sensitivities}. It changes gas rotation while preserving the capture
criterion and all numerical settings.

\textbf{Deposition accounting.} The ensemble calculation tracks
first-entry samples, size-resolved outcomes, conditional transitions,
and radial deposition kernels. For the primary calculation and the Keplerian-gas control,
each accepted integration interval is divided into eight subintervals
using dense output. Ablated mass is placed at each subinterval midpoint; surviving parent mass
is placed at the disk-capture endpoint. Inner-boundary absorption contributes
to Saturn absorption but not to any radial deposition bin. The plotted cumulative profiles use
1-\(R_S\) bins, without a fitted radial redistribution law. The same
orbit integration scheme and deposition sampling are used for the
Keplerian control.
Every mass element is counted once; an encounter that misses the disk
contributes to the encounter denominator but no deposited mass.

\subsection{Radial composition, mass accounting, and numerical
checks}\label{a.9.-radial-composition-mass-accounting-and-numerical-checks}

\textbf{Mass conservation for each trajectory.} Let \(A_v\), \(A_m\),
and \(C\) be the vapor, retained shed-melt, and disk-captured parent
fractions of initial mass. Let \(C_S\) be the surviving fraction
absorbed at \(2R_S\), and \(M_{\rm exit}\) the surviving fraction
crossing the outer exit boundary. These are normalized fractions, not
dimensional masses. The required balance is

\begin{equation}
A_v+A_m+C+C_S+M_{\rm exit}=1.
\label{eq:legacy-A15}
\end{equation}

The total surviving mass retained by Saturn is \(C+C_S\); only \(C\) is credited
to disk deposition. Previously removed vapor and melt are not counted
again as parent mass. Complete ablation leaves no surviving parent.
For the primary population, the largest individual mass residual is below
\(2.8\times10^{-9}\) of initial mass; the weighted budgets close to
within \(1.4\times10^{-12}\). The maximum Jacobi-work residual is
reported in Appendix~\ref{a.8.-orbital-integration-and-conditional-probability-checks}. All disk-capture endpoints lie within
\(R_H\) with a Jacobi constant above the Hill escape-neck value
\(3GM_S/R_H\).

\textbf{Radial mass accounting.} Let \(D_a\) denote the deposited mass
in annulus \(a\) for one trajectory, divided by that trajectory's
initial mass. The radial ledger must satisfy

\begin{equation}
\sum_a D_a=A_v+A_m+C.
\label{eq:legacy-A16}
\end{equation}

The radial composition calculation sums component masses before taking
each ratio in equation~\eqref{eq:legacy-22}. Number-based return probabilities do not
enter that equation. We verify that the mean of the four replica kernels
equals the pooled kernel, that rebinning preserves the radial component
sums, and that the mass-weighted mean of the annular ice fractions
recovers the whole-disk deposited ice fraction. An unweighted mean need
not recover that value.

\textbf{Source-sampling ranges.} Shaded ranges in Figure~\ref{fig:orbital-composition} are the minimum and maximum of four
independently sampled composition profiles. They show the variation from finite source sampling under the adopted
incoming distribution.

\textbf{Spatial-assignment refinement.} This test uses 48 fixed
successful entry geometries from the source ensemble and 24 logarithmic
size bins for each material, giving 3456 size-material histories. We
first compare the production ledger's eight dense-output deposition
subintervals per accepted step with sixteen subintervals. We then repeat
the latter calculation at a relative tolerance of \(2\times10^{-12}\),
half the maximum step, and thirty-two boundary subdivisions. All three
runs use pressure-supported gas. Using identical histories separates the effects of spatial assignment
and orbit integration.

The deposited-mass ledger closes to within \(2.9\times10^{-13}\).
Doubling the deposition subdivisions changes an \(8R_S\)-bin ice
fraction by less than 0.00021; subsequent orbit and event refinement
changes it by less than 0.00040. The corresponding changes in the
inner \(2\)--\(8R_S\) bin are below \(4.8\times10^{-9}\) and
\(3.6\times10^{-7}\). Endpoint classifications and passage counts are
unchanged. Individual first-pass or total delivery fractions change by
less than \(5.8\times10^{-8}\); the largest component-channel change
is \(1.1\times10^{-6}\), and the largest capture-radius shift is
\(0.0015R_S\). These checks establish the numerical accuracy of the displayed bins,
which combine the \(1R_S\) mass-assignment grid, for the adopted gas
domain and capture prescription. Appendix~\ref{a.11.-vertical-extent} separately tests the vertical cutoff; the
physical assignment of captured-parent mass is discussed in Section~\ref{disk-capture-and-saturn-absorption}.

\textbf{Captured-parent endpoint velocities.} Binding, cylindrical and
vertical containment, and the Hill escape-neck barrier are checked at
every disk-capture endpoint. There are 66 retrograde endpoints among
30,405 disk captures. For \(p=3.5\), weighting by the surviving parent
mass credited at capture gives median gas-relative speeds of
\(0.90c_s\), \(0.95c_s\), and \(0.94c_s\) for ice, vapor-only rock,
and rock with retained melt, respectively. The fractions of those
captured-parent masses above \(c_s\) are 0.25, 0.28, and 0.27;
retrograde parents carry 0.023, 0.016, and 0.023. Each material is
normalized to its own surviving disk-captured parent mass, excluding
previously removed vapor and melt and inner-boundary absorption.
These velocities quantify the orbital motion remaining at disk capture.

\subsection{Keplerian numerical control}\label{a.10.-gas-rotation-and-physical-boundary-sensitivities}

\begin{figure}[ht!]
\centering
\includegraphics[width=\textwidth,alt={Four panels compare assigned ice fractions versus distance from Saturn for pressure-supported gas, shown by solid curves, and the Keplerian gas control, shown by dashed curves. Columns distinguish the two rock treatments. Upper panels use p=3.5 with shading spanning four source scrambles; lower panels compare p=3, 3.5, and 4. Both calculations use the same incoming population, radial sampling, and capture criterion. The profiles show outer ice enrichment. Vertical guides mark the present orbits of Titan and Iapetus.}]{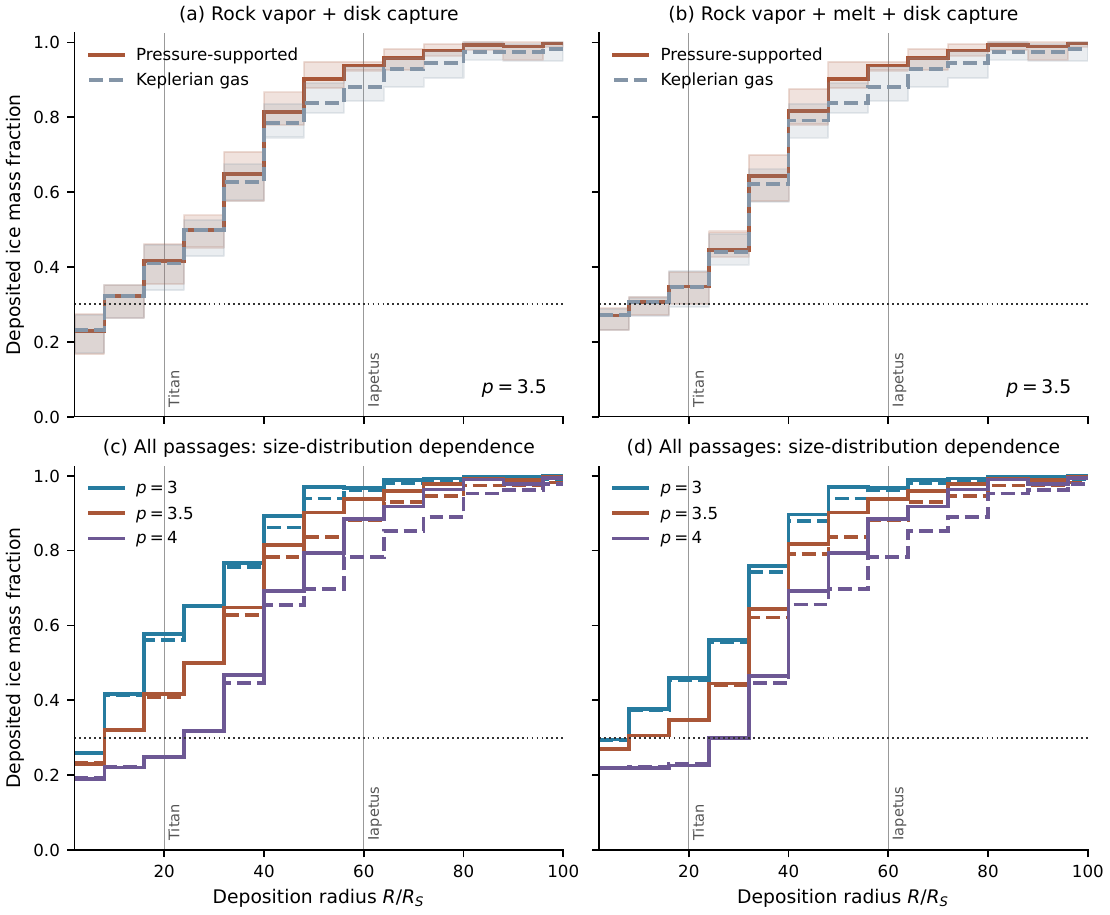}
\caption{Radial ice-fraction comparison in the format of Figure~\ref{fig:orbital-composition},
with pressure-supported gas (solid) and the Keplerian-gas numerical
control (dashed). All curves include all passages and use a 30/70
ice--rock source. Left and right columns give the vapor-only and
retained-melt rock treatments, respectively, including captured parent
mass. Panels (a)--(b) use \(p=3.5\); shading spans the four independent
source scrambles for each gas law. Panels (c)--(d) compare
\(p=3,3.5,4\), with the same solid/dashed distinction at each index.
Radial bins, axes, and source-reference line match Figure~\ref{fig:orbital-composition}. Both
calculations use the same incoming population, bound and disk-contained
capture criterion, and eight dense-output deposition subintervals per
accepted step. Vertical lines mark Titan's and Iapetus's present orbits.\label{fig:gas-control-composition}}
\end{figure}

The main calculation uses the pressure-supported gas law of Section~\ref{orbital-continuation-and-repeated-disk-passages}.
For the control alone we remove its radial pressure term and set

\begin{equation}
v_{g,\phi}=\left(\frac{GM_S}{R}\right)^{1/2}.
\label{eq:legacy-A17}
\end{equation}

Suppressing the headwind isolates its effect in this numerical control.
Pressure-supported
gas lags the circular speed by about \(95\ \mathrm{m\,s^{-1}}\) at \(20R_S\) and
\(1.8\times10^2\ \mathrm{m\,s^{-1}}\) at \(60R_S\), or 0.16 and
0.31 of the local sound speed.

The control repeats all 81,024 histories from the same 512 incoming
geometries, 64 size bins, and three materials. Other assumptions,
boundaries, tolerances, the capture criterion in equation~\eqref{eq:legacy-12e},
and the eight deposition subintervals per step are unchanged. Entry
probability is identical because incoming trajectories are ballistic. Table~\ref{tab:keplerian} records the Keplerian values;
Table~\ref{tab:4} and Figures~\ref{fig:multipass}--\ref{fig:orbital-composition} use pressure-supported gas.

\begin{table}[ht!]
\centering
\caption{Keplerian numerical control for \(p=3.5\). The first two
probabilities are return conditional on first exit and eventual
\(\mathsf K\) (disk capture or Saturn absorption) conditional on first
return, respectively. Delivery is normalized to first-crosser component mass.
The final column is the whole-disk ice fraction when that rock treatment
is combined with ice in the 30/70 source ratio.\label{tab:keplerian}}
\small
\begin{tabular}{lcccc}
\toprule
Treatment & Return & Eventual \(\mathsf K\) & Disk delivery & Ice fraction\\
\midrule
Ice & 0.82 & 0.98 & 0.33 & ---\\
Rock: vapor-only & 0.70 & 0.97 & 0.17 & 0.45\\
Rock: retained melt & 0.70 & 0.95 & 0.18 & 0.44\\
\bottomrule
\end{tabular}
\end{table}

The control has 30,690 disk-capture endpoints, 1,098 inner absorptions,
584 complete ablations, and 48,652 outer exits. These are unweighted
history counts; probabilities and mass efficiencies use the weights
specified in Section~\ref{incoming-ensemble-and-passage-probabilities}. Its maximum individual mass residual is below
\(3.2\times10^{-9}\), and the Jacobi-work residual defined by equation~\eqref{eq:legacy-A12} is below
\(1.2\times10^{-10}\). Both gas laws use the same capture condition, so differences
arise from drag and heating before that endpoint.

Figure~\ref{fig:gas-control-composition} compares the local ice fractions
with identical radial sampling and parent-mass assignment. In
\(16\)--\(24R_S\), the Keplerian control gives 0.41 for vapor-only rock
and 0.35 with retained melt, compared with 0.42 and 0.35 for
pressure-supported gas. In \(56\)--\(64R_S\), both rock treatments give
0.88 in the control and 0.94 with pressure support. Both gas laws
produce an outward gradient in the delivered ice fraction for each of the three size-distribution indices.

\subsection{Vertical-extent sensitivity}\label{a.11.-vertical-extent}

We extend the gas to \(\pm2H(R)\) for the same 512 Hill-boundary states,
64 size bins, three materials, and source weights. The Gaussian density
profile and normalization, temperature, pressure support, radial
boundaries, tolerances, and deposition sampling are unchanged. The wider
domain contains 0.95 of the full Gaussian column instead of 0.68.

The capture prescription still requires binding and containment within the
modeled gas domain. Defining the dimensionless vertical half-extent
\(b_z\), we use

\begin{equation}
\begin{aligned}
|z|&\leq b_z H(R)\quad\text{(gas domain)},\\
r_a\sin i_{\rm orb}&<b_z H(R_{\rm lb})\quad\text{(capture)},\\
b_z&=1\ \text{or}\ 2.
\end{aligned}
\label{eq:legacy-A18}
\end{equation}

Binding and radial conditions in equation~\eqref{eq:legacy-12e} are unchanged;
this tests the additional gas column and wider containment domain together.
Recomputing first entry from the same Hill states admits 481 first
crossers (0.94 of the source), including all 422 reference first crossers
(0.82). The 59 additional entrants follow from the wider boundary, giving
92,352 gas-interacting histories (Table~\ref{tab:vertical-extent}).

\begin{figure}[t!]
\centering
\includegraphics[width=\textwidth,alt={Four panels compare assigned ice fractions versus distance from Saturn with gas extending one scale height above and below the midplane, shown by solid curves, or two scale heights, shown by dashed curves. Both cases use pressure-supported gas and the same incoming population. Columns distinguish the two rock treatments. Upper panels use p=3.5 with shading spanning four source scrambles; lower panels compare p=3, 3.5, and 4. Outer ice enrichment persists at both vertical extents. Vertical guides mark the present orbits of Titan and Iapetus.}]{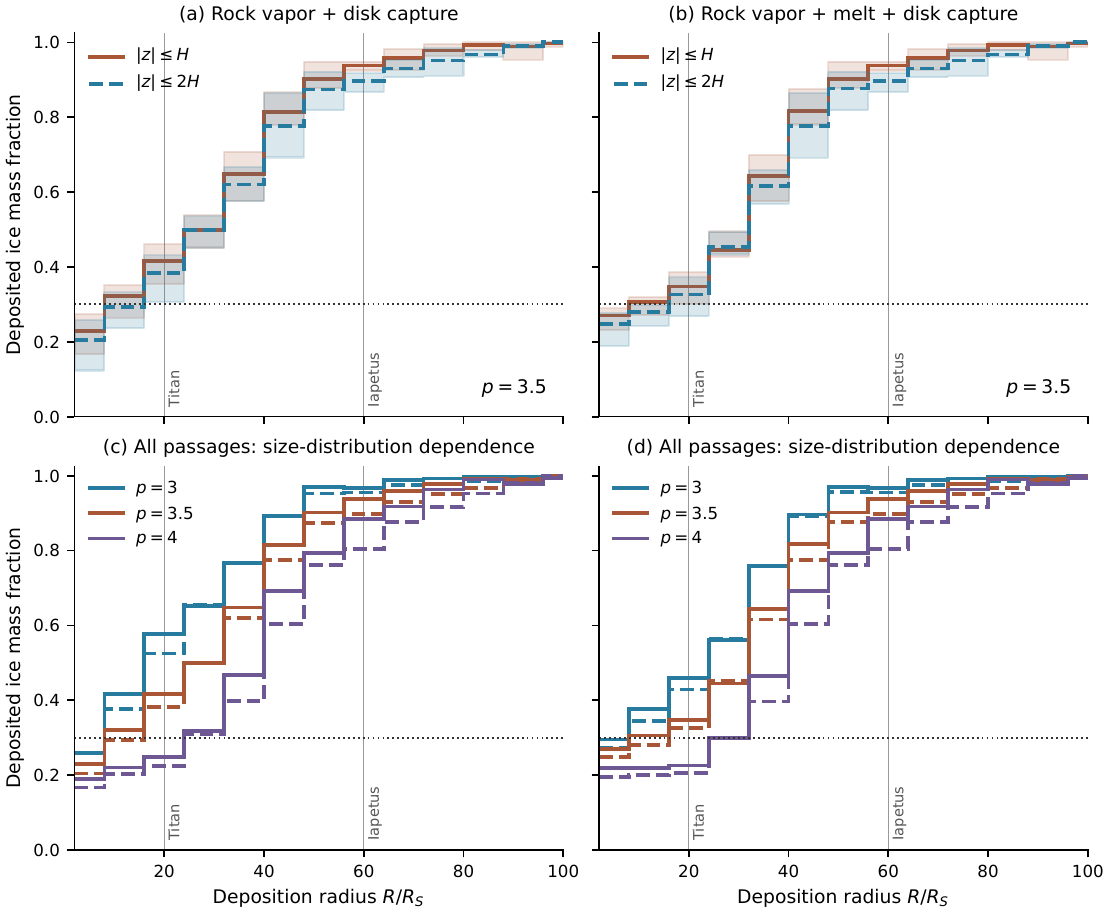}
\caption{Sensitivity of the local ice fraction to gas extending to
\(\pm H\) (solid) or \(\pm2H\) (dashed), with pressure-supported gas
in both cases. The layout, radial bins, and 30/70 ice--rock source match
Figure~\ref{fig:orbital-composition}. Left and right columns give the vapor-only and retained-melt
rock treatments, including captured parent mass. All curves include
all passages. Panels (a)--(b) use \(p=3.5\); shading spans the same
four source scrambles for each extent. Panels (c)--(d) show
\(p=3,3.5,4\), with solid/dashed curves at each index. The gas and containment boundaries vary together as in
equation~\eqref{eq:legacy-A18}. Vertical guides mark the current satellite orbits.
\label{fig:vertical-extent-composition}}
\end{figure}

Figure~\ref{fig:vertical-extent-composition} compares the local
ice fractions using the same radial bins. At \(16\)--\(24R_S\),
the \(H\) values of 0.42 and 0.35 become 0.38 and
0.33 at \(2H\), for vapor-only and retained-melt rock,
respectively. At \(56\)--\(64R_S\), both rock treatments change from
0.94 at \(H\) to 0.90 at \(2H\). Both vertical extents retain the outward gradient in the delivered material for all three
size-distribution indices. At \(p=4\), neither extent enriches the \(16\)--\(24R_S\)
inventory above the source ice fraction. At \(p=3.5\) and \(2H\),
the \(8\)--\(16R_S\) bin is also below the source, with ice fractions
of 0.29 for vapor-only rock and 0.28 with retained melt.

All 92,352 \(2H\) histories reach physical endpoints: 34,301 disk
captures, 1,415 inner absorptions, 512 complete ablations,
and 56,124 outer exits. All disk-captured orbits satisfy the binding
and geometric conditions and lie above the Hill escape-neck Jacobi
barrier. The maximum individual mass residual is below \(7.5\times10^{-9}\),
and the Jacobi-work residual of equation~\eqref{eq:legacy-A12} is below \(7.1\times10^{-11}\).
Refining first-entry detection for all 512 launches changes no
classification. A separate test uses 24 entry geometries (12 common,
12 newly admitted), 24 sizes, and three materials. Doubling deposition
subdivisions changes a plotted-bin ice fraction by less than 0.0011;
tightening the orbit tolerance tenfold, halving the maximum step, and
doubling boundary subdivisions change it by less than 0.00081.
Endpoint classifications agree. The largest disk-capture-radius shift
is \(0.0038R_S\); one trajectory changes from 60 to 61 gas visits with
the same disk-capture outcome.

\begin{table}[ht!]
\centering
\caption{Matched vertical-extent test at \(p=3.5\). Each entry gives
\(H\) / \(2H\). First \(\mathsf K\) is conditional on first crossing;
return is conditional on first exit; eventual \(\mathsf K\) is conditional
on first return, with the definitions in Table~\ref{tab:4}. \(\mathsf K\) includes
disk capture and Saturn absorption. Disk delivery includes
vapor, retained melt where applicable, and surviving disk parents,
normalized to the common incoming component mass; inner-absorbed mass
is excluded.\label{tab:vertical-extent}}
\small
\begin{tabular}{lcccc}
\toprule
Treatment & First \(\mathsf K\) & Return & Eventual \(\mathsf K\) & Disk delivery\\
\midrule
Ice & 0.54 / 0.57 & 0.82 / 0.71 & 0.98 / 0.95 & 0.27 / 0.30\\
Rock: vapor-only & 0.25 / 0.26 & 0.70 / 0.67 & 0.97 / 0.94 & 0.14 / 0.16\\
Rock: retained melt & 0.24 / 0.25 & 0.70 / 0.67 & 0.95 / 0.93 & 0.15 / 0.17\\
\bottomrule
\end{tabular}
\end{table}

\subsection{Jupiter implementation and numerical checks}\label{a.jupiter-checks}

The Jupiter calculation uses
\(R_J=7.15\times10^7\) m,
\(GM_J=1.267\times10^{17}\ \mathrm{m^3\,s^{-2}}\), and a
circular heliocentric orbit of 5.20 au. These numerical constants
retain additional digits for reproducibility. The Hill radius is
\(742.80R_J\). The units, Hill equations, drag-work integral,
capture geometry, material laws, and mass-ledger construction follow
Appendices~\ref{a.8.-orbital-integration-and-conditional-probability-checks}
and~\ref{a.9.-radial-composition-mass-accounting-and-numerical-checks},
with Jupiter's mass and radius replacing Saturn's.

The same four Sobol seeds, 731--734, each generate 128 source samples.
There are 413 first crossers, giving 79,296 gas-interacting histories
for 64 sizes and three materials. First-entry detection uses 32
dense-output subdivisions per step, relative tolerance
\(2\times10^{-10}\), and the same maximum-step prescription as
Saturn. Tightening this tolerance to \(2\times10^{-12}\), halving
the maximum step, and increasing to 64 subdivisions change no
first-entry classification among the 512 launches. Subsequent
gas-interacting trajectories use the Saturn production settings:
relative tolerance \(2\times10^{-11}\), step fraction 0.05,
16 boundary/capture scans, and eight radial-deposition subdivisions.

All 79,296 histories reach physical endpoints: 20,893 disk captures,
361 planetary absorptions, 3,191 complete ablations, and 54,851 Hill
exits. The other 99 source encounters comprise 94 ballistic Hill
exits and five direct planetary absorptions. Every disk-capture
endpoint satisfies the binding and containment conditions and lies
above the Hill escape-neck Jacobi barrier. Individual mass residuals
are below \(2.0\times10^{-8}\), the dimensionless Jacobi-work
residual is below \(1.2\times10^{-10}\), and weighted mass budgets
close to \(3.2\times10^{-12}\). Rebinning preserves the radial
component sums, and the four replica kernels average to the pooled
kernel.

A paired refinement uses 16 fixed entry geometries, 24 sizes, and
three materials, or 1,152 histories per setting. Doubling the radial
deposition subdivisions changes a \(4R_J\)-bin ice fraction by less
than 0.0040. Subsequently tightening the orbital tolerance tenfold,
halving the maximum step, and doubling boundary scans change it by
less than 0.0016. Endpoint classifications agree; the largest
individual channel change is \(1.2\times10^{-7}\), and the largest
capture-radius shift is \(5.5\times10^{-5}R_J\).

Doubling the size grid from 64 to 128 bins for all 512 source
states gives 158,592 histories, with identical entry states and all
other numerical settings unchanged. Across the three size-distribution
indices and both rock treatments, the largest whole-disk ice-fraction
change is 0.00065. The largest change in a displayed \(4R_J\) bin is
0.011 (0.0067 at \(p=3.5\)); over the full \(2\)--\(100R_J\)
domain it is 0.028. The quoted \(p=3.5\) compositions in
\(14\)--\(18R_J\) and \(26\)--\(30R_J\) are unchanged at two
significant figures. Both size grids retain the outward increase
across the Galilean region.

With drag and thermal recession disabled, four controlled trajectories
at 6, 15, 26, and \(60R_J\) conserve the dimensionless Hill Jacobi
constant to \(2.2\times10^{-13}\); removing the solar terms also
conserves two-body orbital energy to \(1.3\times10^{-13}\).
An independent, uninterrupted event integration of five controlled
ice, rock, and retained-melt histories reproduces the endpoints, with
individual mass channels differing by less than \(3.5\times10^{-10}\).
The planet extension reproduces Saturn's environment on a 501-point
radial grid and six complete Saturn trajectories exactly; the Saturn
production data are unchanged.

\subsection{Jovian focusing-parameter comparison}\label{a.jupiter-focusing}

We repeat the Jovian ensemble at \(\Theta=100\) and 25, using the
same 512 Sobol coordinates, 64 size bins, material treatments,
pressure-supported disk, and capture prescription. Each coordinate
set is mapped to the focused cross section appropriate to its
\(\Theta\); thus the velocities and first-crossing states change
with the chosen incoming population. The asymptotic encounter speed
is 4.2 or \(8.4\ \mathrm{km\,s^{-1}}\), respectively. The gas-relative
speed along each trajectory determines its drag and heating.

Both members of this comparison use 64 dense-output subdivisions per
ballistic entry step. This resolves two grazing entries missed in an
initial \(\Theta=25\) calculation with 32 subdivisions. Increasing
to 128 subdivisions, tightening the entry tolerance to
\(2\times10^{-12}\), and halving the maximum step preserve all
512 first-entry classifications at each focusing parameter. The
radial deposition sampling and subsequent trajectory settings are
identical to those in Appendix~\ref{a.jupiter-checks}.

At \(\Theta=25\), 393 first crossers generate 75,456 histories:
11,171 disk captures, 149 planetary absorptions, 2,077 complete
ablations, and 62,059 Hill exits. The remaining source encounters
are 116 ballistic Hill exits and three direct planetary absorptions.
Every disk-capture endpoint satisfies the binding and containment
criteria and lies above the Hill escape-neck Jacobi barrier. Individual
mass residuals are below \(1.1\times10^{-8}\), Jacobi-work residuals
are below \(1.3\times10^{-10}\), and the largest weighted mass-budget
residual is \(4.0\times10^{-12}\).

The matched \(\Theta=100\) run retains the 413 first crossers and
79,296 histories of Appendix~\ref{a.jupiter-checks}, with identical
endpoint counts. Increasing its entry sampling from 32 to 64
subdivisions changes the compared ice fractions by less than 0.000077;
the quoted bulk, Ganymede-region, and Callisto-region values at
\(p=3.5\) are unchanged at two significant figures.
Table~\ref{tab:jupiter-focusing} compares the matched runs. At
\(\Theta=25\) and \(p=3.5\), the four source scrambles give whole-disk
fractions of 0.56--0.61 and 0.51--0.55 for the two rock treatments.
The corresponding ranges are 0.37--0.43 and 0.30--0.35 in
\(14\)--\(18R_J\), and 0.49--0.58 and 0.39--0.44 in
\(26\)--\(30R_J\). These ranges describe source sampling.

\begin{table}[t!]
\centering
\caption{Jovian ice fractions for matched focusing-parameter runs.
Each pair gives \(\Theta=100/25\), with source ice fraction 0.30.
The inventory includes delivered water vapor, retained melt where
specified, and disk-captured remnants from all passages; planetary
absorption is excluded. Columns give ratios of summed component mass
within the indicated radial intervals.}
\label{tab:jupiter-focusing}
\begin{tabular}{llccc}
\toprule
Rock treatment & \(p\) & \(2\)--\(100R_J\) & \(14\)--\(18R_J\) & \(26\)--\(30R_J\)\\
\midrule
Rock: vapor loss only & 3 & 0.51/0.60 & 0.52/0.50 & 0.70/0.71\\
 & 3.5 & 0.49/0.59 & 0.37/0.41 & 0.54/0.56\\
 & 4 & 0.46/0.57 & 0.22/0.30 & 0.38/0.39\\
Rock: vapor loss and retained melt & 3 & 0.45/0.50 & 0.41/0.37 & 0.57/0.52\\
 & 3.5 & 0.45/0.53 & 0.33/0.33 & 0.46/0.43\\
 & 4 & 0.45/0.55 & 0.22/0.28 & 0.34/0.33\\
\bottomrule
\end{tabular}
\end{table}

For \(\Theta=25\), numerical refinement uses 16 fixed entry
geometries, 24 sizes, and three materials, or 1,152 histories at each
setting. Doubling radial-deposition subdivisions changes the displayed
\(4R_J\)-bin fractions by less than 0.0043. Tightening the orbital
tolerance tenfold, halving the maximum step, and doubling boundary
scans then change them by less than 0.0013. All endpoint
classifications agree. Individual mass-channel changes are below
\(2.0\times10^{-7}\), and capture-radius shifts are below
\(7.5\times10^{-6}R_J\). The default launch and first-entry states
for 32 Sobol samples at each planet are identical before and after
adding the selectable focusing parameter.

\clearpage
% Keep normal reference line spacing when balancing the final two columns.
\makeatletter
\let\raggedcolumn@sw\@firstoftwo
\makeatother
\bibliographystyle{aasjournalv7.1}
\bibliography{references}

\end{document}